\documentclass[doublecol]{epl2} 
\graphicspath{{pics/}}
\usepackage{amsmath}
\usepackage{amsfonts}
\usepackage{graphicx}

\title{Route to hyperchaos in Rayleigh-B{\'e}nard convection}
\author{R. Chertovskih\inst{1} \and E.V. Chimanski \inst{1} \and 
E.L. Rempel\inst{1,2}}
\shortauthor{R. Chertovskih \etal}

\institute{
  \inst{1} Aeronautics Institute of Technology  -- S\~ao Jos\'e dos Campos, 
  S\~ao Paulo 12228-900, Brazil\\
  \inst{2} National Institute for Space Research and World Institute 
  for Space Environment Research --  
  P.O. Box 515, S\~ao Jos\'e dos Campos, S\~ao Paulo 12227-010, Brazil
}

\abstract{Transition to hyperchaotic regimes in Rayleigh-B\'enard
convection  in a square periodicity cell is studied by three-dimensional
numerical simulations. By fixing the Prandtl number at $P=0.3$ and varying the
Rayleigh number as a control parameter, a bifurcation diagram is
constructed where a route to hyperchaos involving quasiperiodic regimes with two and
three incommensurate frequencies, multistability, chaotic intermittent
attractors and a sequence of boundary and interior crises is shown.
The three largest Lyapunov exponents exhibit a linear scaling with the
Rayleigh number and are positive in the final hyperchaotic attractor.
Thus, a route to weak turbulence is found.}

\begin{document}
\maketitle

\section{Introduction}

Thermal convection refers to the motion of a heat-conducting fluid due to the
presence of temperature differences.  Convective flows are of interest
in many areas ranging from technological processes (cooling of
electronic devices, drying, material processing) to
natural phenomena (convection in the terrestrial atmosphere, oceans, mantle,
outer core and stellar convection).  Origin of magnetic fields of many
astrophysical objects is explained in the framework of the dynamo theory
\cite{moffat}, where magnetic fields are generated by convective
motions of electrically conducting fluids in their interior; chaotic
convective flows play an important role in the fast dynamo theory.

Thermal convection in the dimensionless form is characterised by the Rayleigh number (Ra),
that measures the magnitude of thermal buoyancy force, and the Prandtl
number ($P$), the ratio of kinematic viscosity to thermal diffusivity.
In the context of Astrophysics, values of the Prandtl number vary from small
(for the solar convective zone, $P\sim 10^{-7}$ \cite{ossen03}) to large
(for mantle convection, $P\sim10^{23}$ \cite{schu01}) values. The Prandtl
number for the terrestrial outer core has intermediate values estimated
to be between $0.1$ and $0.5$ \cite{olson07,fearn}, however, in many
terrestrial convective dynamo simulations a larger value ($P\sim1$) is
employed (see, e.g., \cite{jones11} and references therein) using the
argument that it expresses the ``effective'' or ''turbulent'' value of the
diffusivity \cite{gub13}.

In the dynamo theory, interest in the use of realistic values for the
Prandtl number was enhanced by the following findings. Varying the value
of the Prandtl number from 0.1 to 10 was found to have a strong
influence on the morphology and dynamics of convection in the Earth's
outer core \cite{calk12}. Strong dependence of the magnetic fields generated
by convective flows on the value of the Prandtl number was found for
$0.2\le P \le 5 $ \cite{soltis14}.  In \cite{busse00}, moderately low
Prandtl numbers were beneficial for magnetic field
generation in rotating spherical shells and special attention was
devoted to dynamos for $P=0.1$. Analysis of the kinematic dynamo problem
in \cite{podv08} showed that convective attractors for $P=0.3$ are
beneficial for the magnetic field generation (in comparison to the
attractors for $P=1$ and 6.8). 

Most of the convective flows in nature are turbulent. Thermal convection
in a plane horizontal layer, called Rayleigh-B\'enard convection, has
been used for decades as one of the simplest examples of
realistic hydrodynamic systems driven out of equilibrium where the
simple flow becomes complex (turbulent) in a variety of
bifurcation sequences, revealing different mechanisms of instability and
demonstrating many common nonlinear dynamics phenomena, e.g.,
spontaneous symmetry breaking, pattern formation, intermittency,
synchronisation, etc. (for a review, see \cite{boden,cross}). 
A common approach to study transition to
turbulence in the framework of the dynamical systems theory 
is to study the evolution from simple non-chaotic (steady, time-periodic
and quasiperiodic) attractors of the convective system to the chaotic
ones.  In chaotic attractors, trajectories are sensitive to initial
conditions, i.e., initially close trajectories diverge in time, which is
quantitatively characterised by the Lyapunov spectrum.  
Turbulent systems usually display hyperchaos, i.e., more than one positive
Lyapunov exponent.

There is a wide range of works on bifurcation analysis of
Rayleigh-B\'enard convection as a function of Ra and $P$, most of which
are devoted to the formation and destabilisation of steady planar convective
rolls \cite{chandra,getling,busbol84,bolbus85} and transition to chaos
\cite{meneguz87,podv08}. In \cite{boronska}, a bifurcation diagram is presented
where a series of steady states representing different patterns is numerically
obtained as a function of Ra in a small cylindrical domain. Also in cylindrical
domains, hyperchaotic states were found in studies of spiral defect chaos
using simulations of three-dimensional Rayleigh-B\'enard convection, where
the spectrum of Lyapunov exponents was used to quantify extensivity
in spatiotemporal chaos \cite{egolf,karimi}.

Still regarding transition to hyperchaos in Rayleigh-B\'enard convection, in
\cite{paul11}, a square periodicity cell with aspect ratio $L=2\sqrt{2}$
and stress-free boundaries was studied for $P=6.8$ in two-dimensional
convection for Ra up to 32875 using a low-dimensional model (16 Fourier
modes); the sequence of regimes is as follows: time-periodic,
quasiperiodic with two basic frequencies, phase-locked (periodic) and
then chaotic state (some of the chaotic states are hyperchaotic with two
positive Lyapunov exponents).

In this letter we study the transition to hyperchaotic regimes 
(also referred to as transition to weak turbulence \cite{man06}) in
three-dimensional Rayleigh-B\'enard convection for $P=0.3$ in a square
convective cell for Ra increasing from 1720 (time-periodic state) to
2500 (hyperchaotic state). We show that a sequence of crises involving
quasiperiodic and chaotic attractors, as well as chaotic saddles, is
responsible for the evolution of the attractors of the system from
periodic convective states to hyperchaos with at least three positive Lyapunov 
exponents.

\section{Statement of the problem and solution} A Newtonian fluid flow
in a horizontal plane layer is considered, where the fluid is uniformly
heated from below and cooled from above. Fluid flow is buoyancy-driven
and the Boussinesq approximation is assumed. 
We adopt the vertical size of the layer as a length scale, the vertical
heat diffusion time as a time scale, and the vertical temperature
gradient as a temperature scale. Then, in a Cartesian reference frame with the
orthonormal basis $({\bf e}_1, {\bf e}_2, {\bf e}_3)$, where ${\bf e}_3$
is opposite to the direction of gravity, the equations governing the
convective system are (see, e.g., \cite{getling}):
\begin{gather}
{\partial {\bf v}}/{\partial t}=
P\nabla^2{\bf v}
+{\bf v}\times(\nabla \times {\bf v})
+P{\rm Ra}\theta{\bf e}_3-\nabla p,
\label{NSlayer} \\
{\partial\theta}/{\partial t}=
\nabla^2\theta
-({\bf v}\cdot\nabla)\theta+v_3,
\label{Tlayer} \\
\nabla\cdot{\bf v}=0,
\label{Vsol}
\end{gather}
where ${\bf v}({\bf x},t)=(v_1,v_2,v_3)$ is the fluid velocity, 
$p({\bf x},t)$ the pressure, and 
$\theta({\bf x},t)=T({\bf x},t)-(T_1+(T_2-T_1)x_3)$ 
is the difference between the temperature $T$ and its 
linear profile. Temperatures of the horizontal boundaries 
at the bottom, $T_1$, and top, $T_2$, are maintained constant, with
$T_1>T_2$. Here, the spatial coordinates 
are ${\bf x}=(x_1,x_2,x_3)$ and $t$ stands for time. 
The non-dimensional parameters are the Prandtl number, 
$P$, and the Rayleigh number, Ra. 

The horizontal boundaries of the layer are assumed to be stress-free,
$\partial v_1/\partial x_3=\partial v_2/\partial x_3=v_3=0$, and
maintained at constant temperatures, $\theta=0$.  A square convective
cell is considered, ${\bf x}\in[0,L]^2\times[0,1]$, and all the fields are
periodic in the horizontal directions, $x_1$ and $x_2$, with period $L$. 
The linear theory \cite{chandra} suggests that at the onset of convection 
in an infinite layer (Ra$_{\rm c}$=657.5
for the boundary conditions under consideration) the critical horizontal
wavenumber is $\pi/\sqrt{2}$, independently of the Prandtl number; here,
$L=4$ is taken, hence the most unstable mode at the onset is aligned with
the diagonal of the cell.

Following \cite{podv08}, we studied attractors of the convective system
for \mbox{$P=0.3$} and $1720\le{\rm Ra}\le2500$. The attractors were
obtained integrating the system in time starting from 
an attractor for a neighboring value of Ra (in most
cases located at distance 10); the first 1500 eddy turnover times of the
largest eddies were disregarded as transients. To check if
multiple attractors co-exist, for several values of Ra the problem was
solved for four initial conditions defined by random Fourier
coefficients of $\bf v$ and $\theta$ with exponentially decaying
spectrum and the following values of kinetic energy:
$E_v(0)=1,\,100,\,400,\,2500$.  

For a given initial condition, the system under consideration is
integrated numerically forward in time using the standard pseudospectral
method \cite{canuto1}: the fields are represented as Fourier series in
all spatial variables (exponentials in the horizontal directions,
sine/cosine in the vertical direction), derivatives are computed in the
Fourier space, multiplications are performed in the physical space, and
Orszag's 2/3-rule is applied for dealiasing. The system of ordinary differential equations for
the Fourier coefficients is solved using the third-order exponential
time-differencing method ETDRK3 \cite{coxmatt} with constant time step
$\Delta t = 5\cdot10^{-4}$. The spatial resolution was chosen to be
$32\times32\times16$ Fourier harmonics (multiplications were performed
on a uniform $48\times48\times24$ grid). For all solutions, the time-averaged
energy spectra of the velocity decay at least by 5 orders of magnitude.
For each branch of attractors, several runs with doubled spatial and
temporal resolution showed no qualitative difference.

The three largest Lyapunov exponents,
$\lambda_1\ge\lambda_2\ge\lambda_3$, were computed using the technique
described in \cite{hram}, with recourse to the operator of linearisation
of the governing equations (\ref{NSlayer})--(\ref{Vsol}) and the
Gramm-Shmidt orthonormalisation. Translational invariance of the
convective system in the horizontal directions gives rise to two
vanishing Lyapunov exponents, which were disregarded in computations by
removing the corresponding components of the perturbations 
(note that for non-steady attractors at least one vanishing 
Lyapunov exponent remains since eqs. (\ref{NSlayer})--(\ref{Vsol}) are an autonomous system).
According to the general theory \cite{ott}, for a stable limit cycle (time-periodic
attractor) $\lambda_1=0,\, \lambda_2,\lambda_3<0$; for a $k$-torus
(quasi-periodic attractor with $k$ basic frequencies) the largest $k$
Lyapunov exponents vanish; chaotic and hyperchaotic attractors are
characterised by at least one and two positive Lyapunov exponents,
respectively.

In our computations we traced the kinetic energy,
$
E_v(t)=\int_0^L\int_0^L \int_0^1
{\bf v}^2{\rm d}{\bf x}/({2L^2}),
$
as well as the Fourier harmonics of the flow velocity, 
$\widehat{\bf v}_{\bf k}(t)=(\widehat{v}_{\bf k}^1, 
\widehat{v}_{\bf k}^2,\widehat{v}_{\bf k}^3)$, for some wave vectors, 
${\bf k}=(k_1,k_2,k_3)$. Poincar\'e sections were constructed for the Fourier
harmonic of the fluid velocity for ${\bf k}=(1,1,1)$ on the quadrant
$(|\widehat{v}^1_{\bf k}|,|\widehat{v}^2_{\bf k}|)$, where intersection
of the trajectory with the plane $|\widehat{v}^3_{\bf k}|=0.25$ was considered.  
Using the solenoidality condition (\ref{Vsol}) for the
Fourier coefficient $\widehat{\bf v}_{1,1,1}$ and the inequalities
$|z_1|-|z_2|\le|z_1+z_2|\le|z_1|+|z_2|,\, \forall
z_1,z_2\in\nolinebreak\mathbb{C}$, one proves that all the points on the
Poincar\'e section  belong to the semi-infinite strip 
$||\widehat{v}^1_{1,1,1}|-|\widehat{v}^2_{1,1,1}| |\le0.5$, 
$|\widehat{v}^1_{1,1,1}|+|\widehat{v}^2_{1,1,1}|\ge0.5$. Absolute
values eliminate drifting frequencies from consideration. On the Poincar\'e
section, a time-periodic attractor and a 2-torus appear as 
a finite set of points and a curve, respectively. 

\section{Results}

\begin{figure}
\centerline{
\includegraphics[scale=0.72]{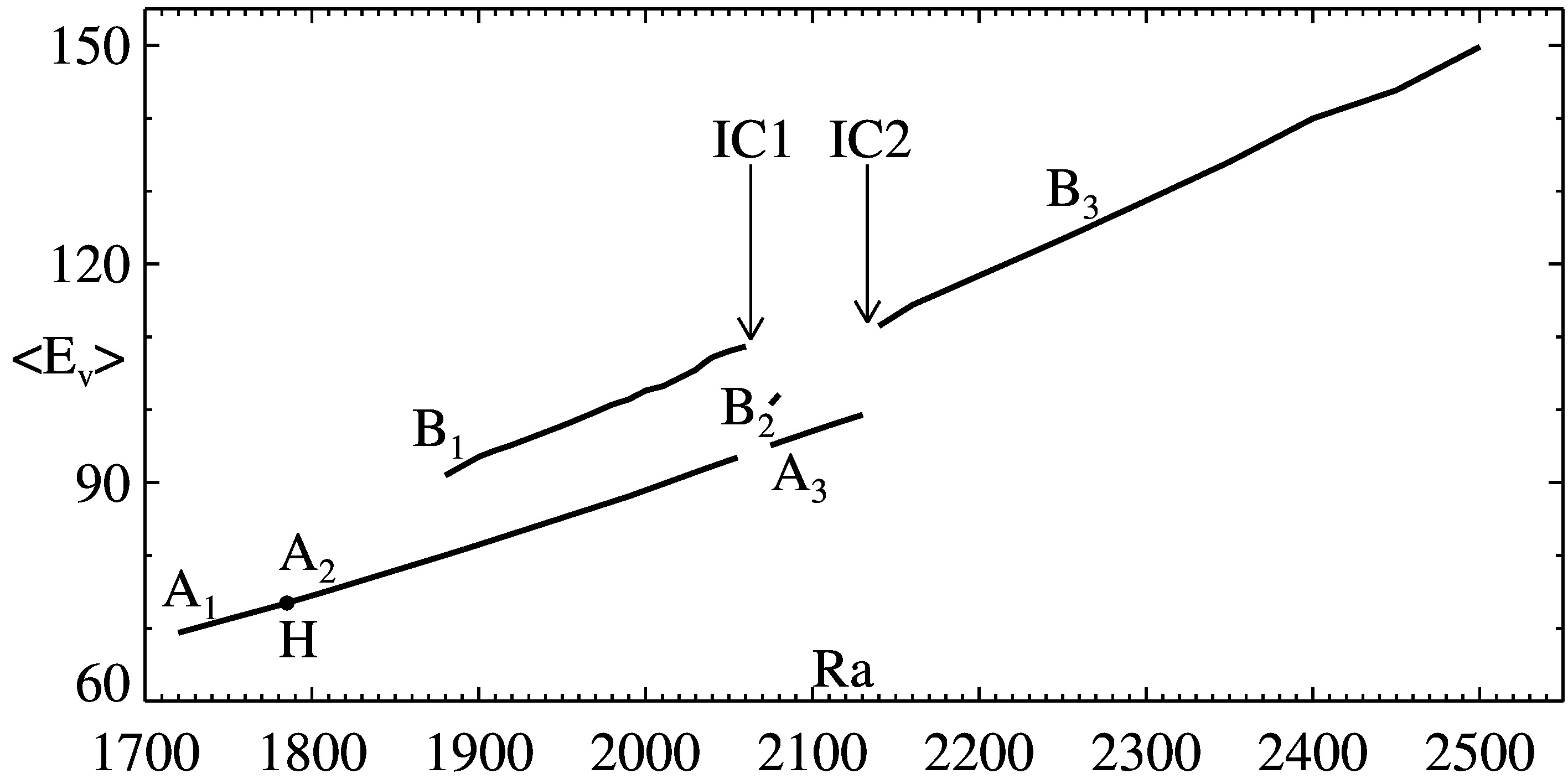}
}
\caption{
Time-averaged kinetic energy versus Rayleigh number of the attractor families. 
IC denotes an interior crisis, H a Hopf bifurcation.
}
\label{fig:en}
\end{figure}

In what follows, attractors of the convective system for 
$1720\le{\rm Ra}\le2500$ are discussed. 
This range of Ra was also considered in \cite{podv08}, where bifurcations of 
the convective attractors were studied for $657.5\le{\rm Ra}\le2500$,
with transition to chaotic
attractors following the sequence: periodic-quasiperiodic-chaotic.
However, no detailed study of the chaotic attractors was performed. In
the present letter, results of a detailed analysis of the transition to chaos are reported.
Branches of attractors found for different intervals of Ra
are shown in fig.~\ref{fig:en}. 

Attractors of the branch $\rm{A}_1$ are periodic states, existing for
$1720\le{\rm Ra}\le1780$, and are symmetric at any time with respect to
rotation about the vertical axis by $\pi/2$.  On increasing Ra, the
symmetry is broken and the branch $\rm{A}_2$ of time-periodic states
(with trivial symmetry group) drifting along the horizontal directions
emanates. Although the convective regimes constituting $\rm{A}_2$ are
formally quasiperiodic, for simplicity, we classify them as periodic
(also referred to as relative periodic orbits \cite{chossat}) since they
are periodic in a co-moving reference frame. In what follows, we ignore
drift frequencies when classifying an attractor as periodic or quasiperiodic.

\begin{figure}[!]
\centerline{
\includegraphics[scale=0.55]{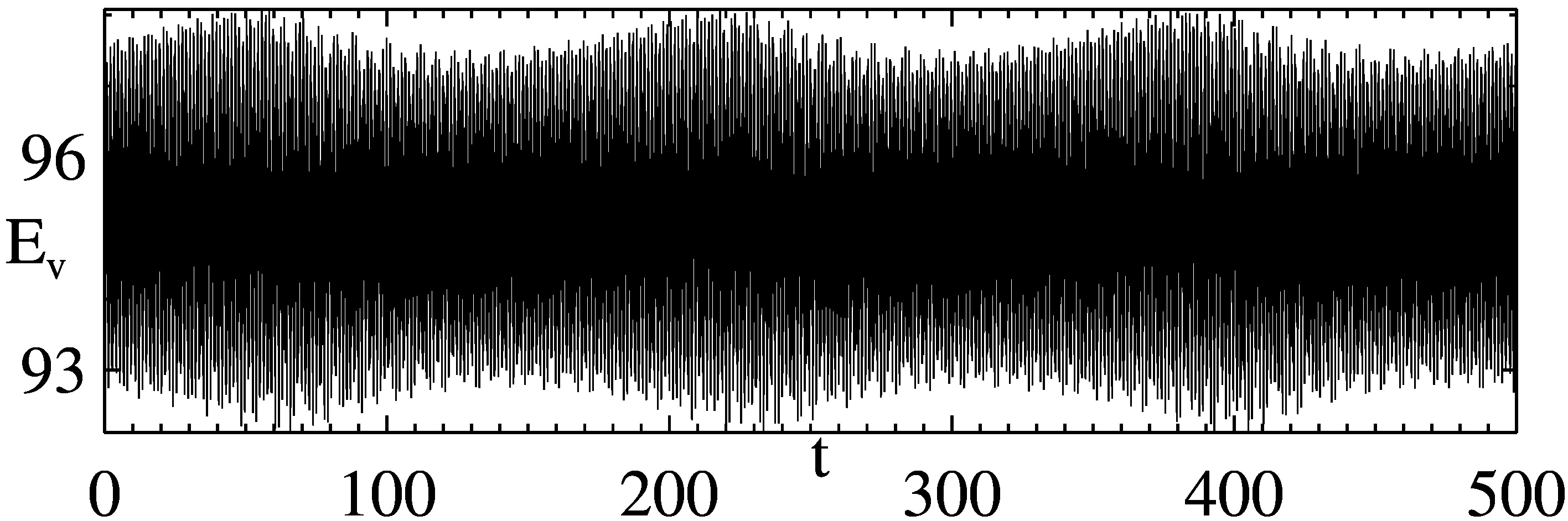}\hspace*{-0.2cm}
\includegraphics[scale=0.55]{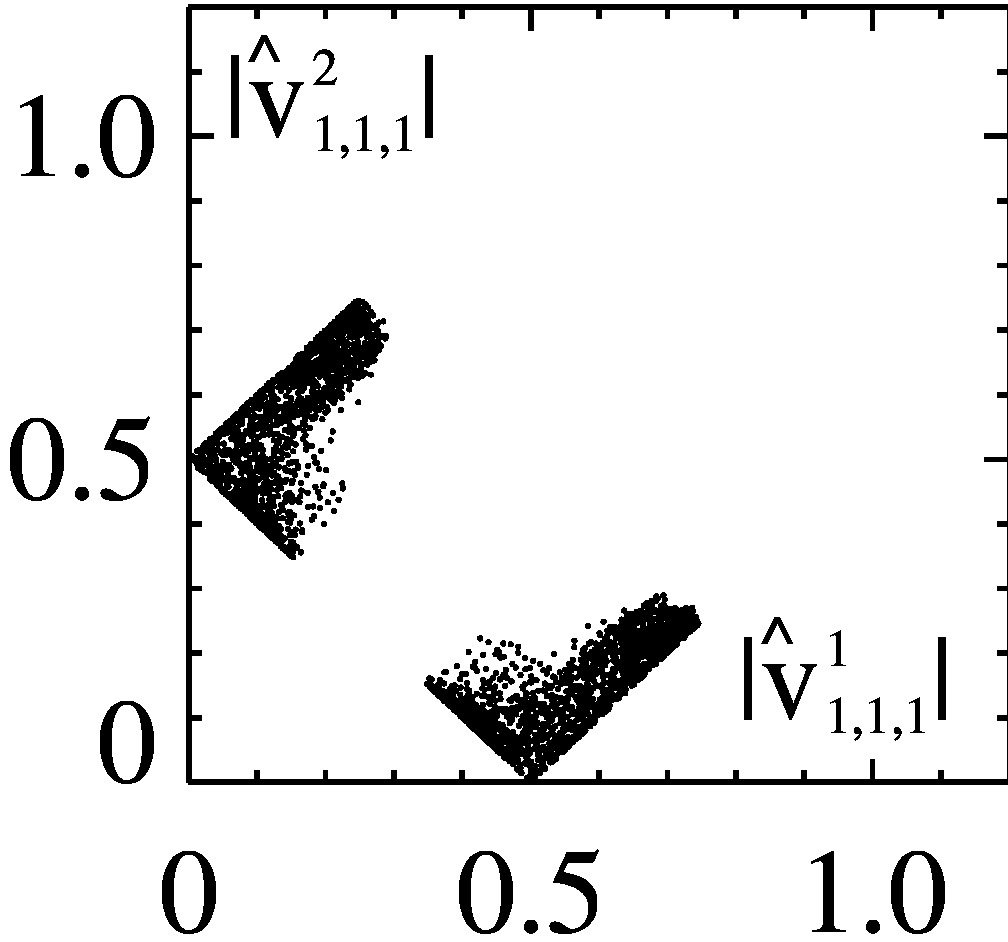}
}
\caption{
Kinetic energy evolution and the Poincar\'e section for the 
quasiperiodic attractor at Ra=2075 from the family~$\rm{A}_3$. 
Left panel: kinetic energy $E_v$ versus time; 
right panel: Poincar\'e section defined by 
$|\widehat{v}^3_{1,1,1}|=0.25$ on the 
$(|\widehat{v}^1_{1,1,1}|,|\widehat{v}^2_{1,1,1}|)$ quadrant. 
}
\label{fig:tu}
\end{figure}

A branch of quasiperiodic attractors, $\rm{A}_3$, exists for
$2075\le{\rm Ra}\le2130$, with three basic frequencies (the Poincar\'e
plane shown on fig.~\ref{fig:tu} confirms that the regime has at least
three incommensurate frequencies). 
We have also checked that for all attractors from
this branch all the three largest Lyapunov exponents vanish.

\begin{figure}[!t]
\centerline{
\includegraphics[scale=0.50]{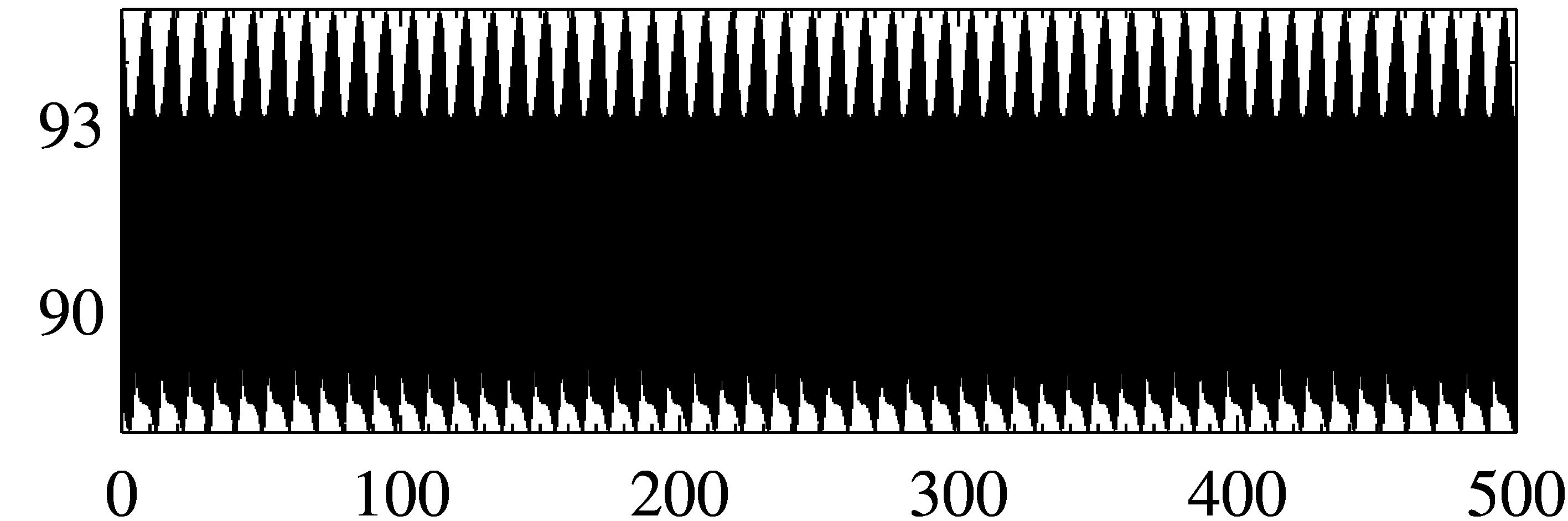}
\includegraphics[scale=0.50]{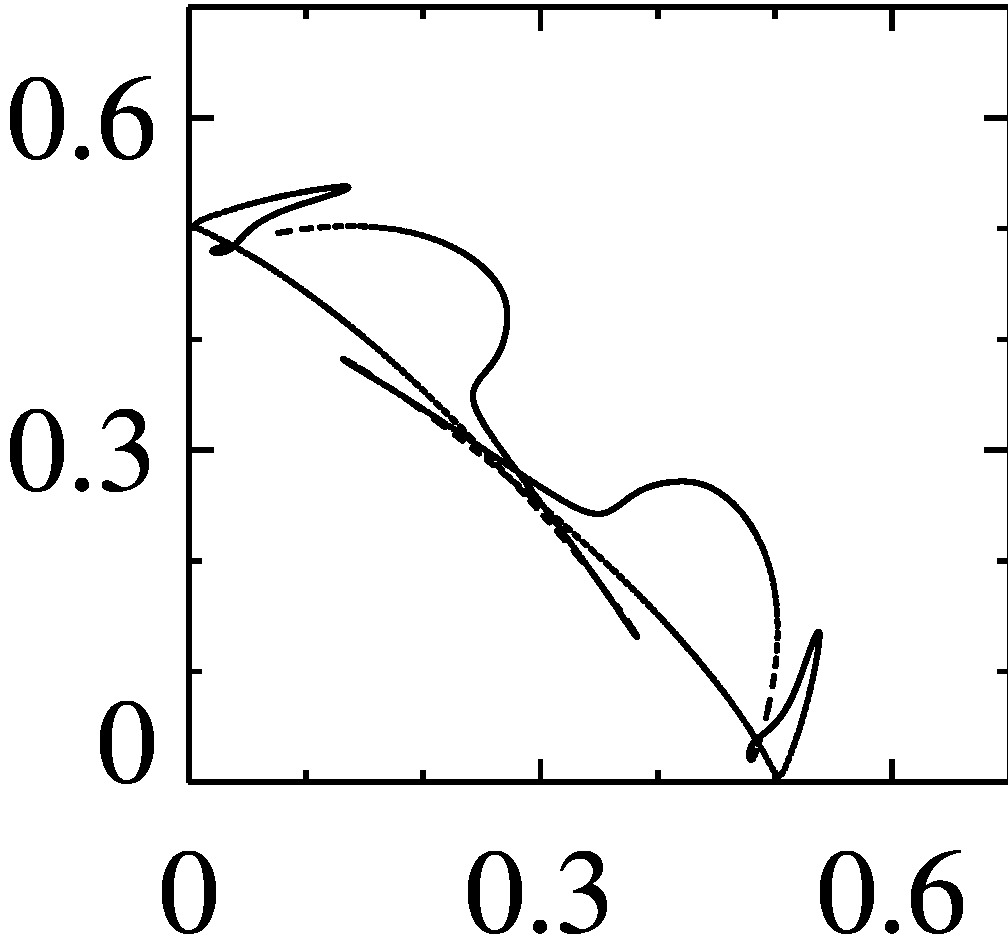}
}
\vspace*{-1.7cm}
\centerline{\hspace*{4.0cm}(a)}
\vspace*{1.2cm}
\centerline{
\includegraphics[scale=0.50]{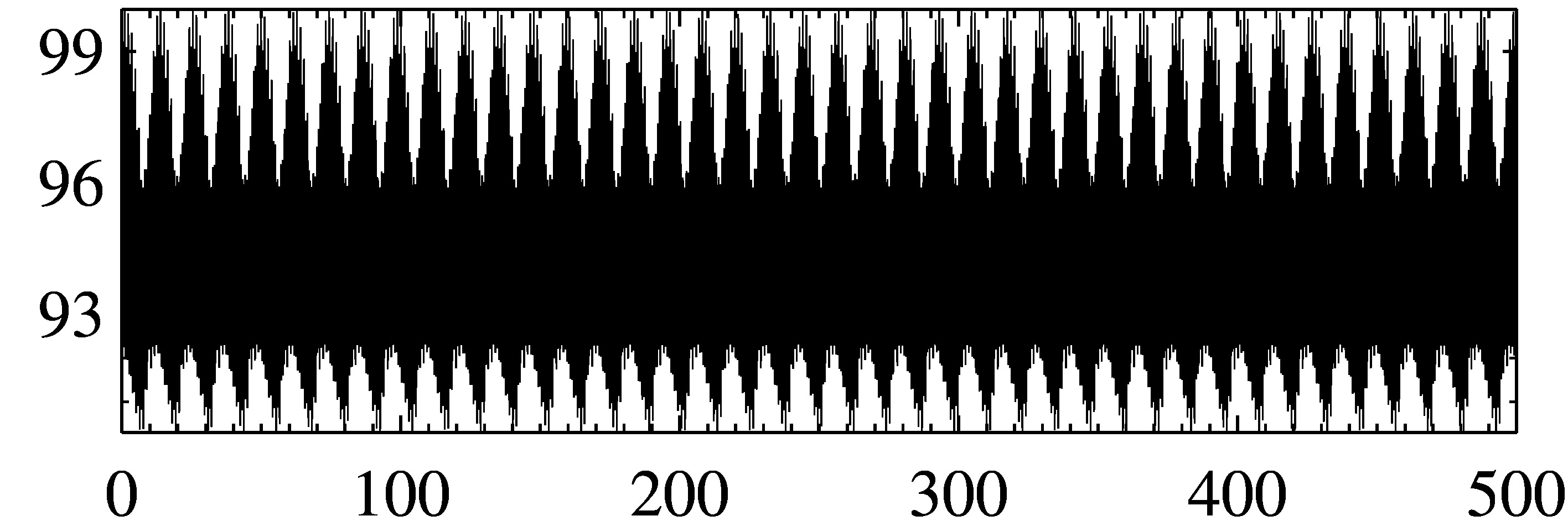}
\includegraphics[scale=0.50]{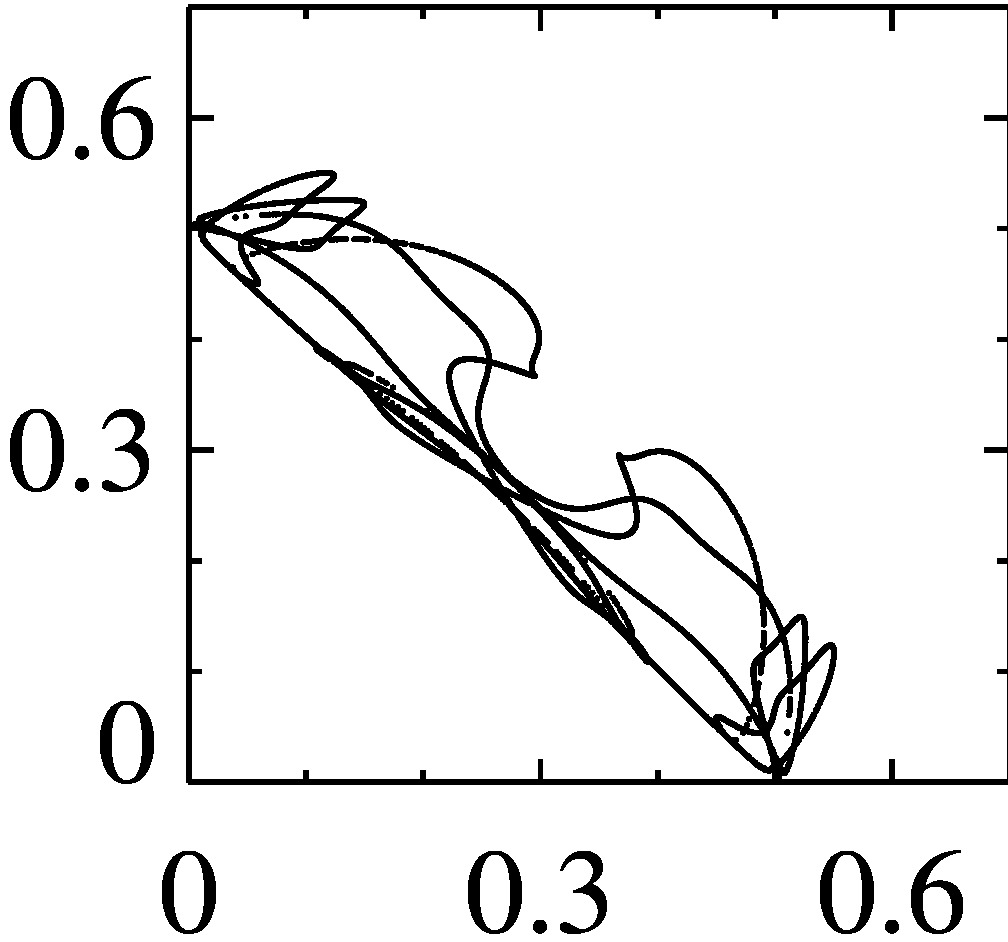}
}
\vspace*{-1.7cm}
\centerline{\hspace*{4.0cm}(b)}
\vspace*{1.2cm}
\centerline{
\includegraphics[scale=0.50]{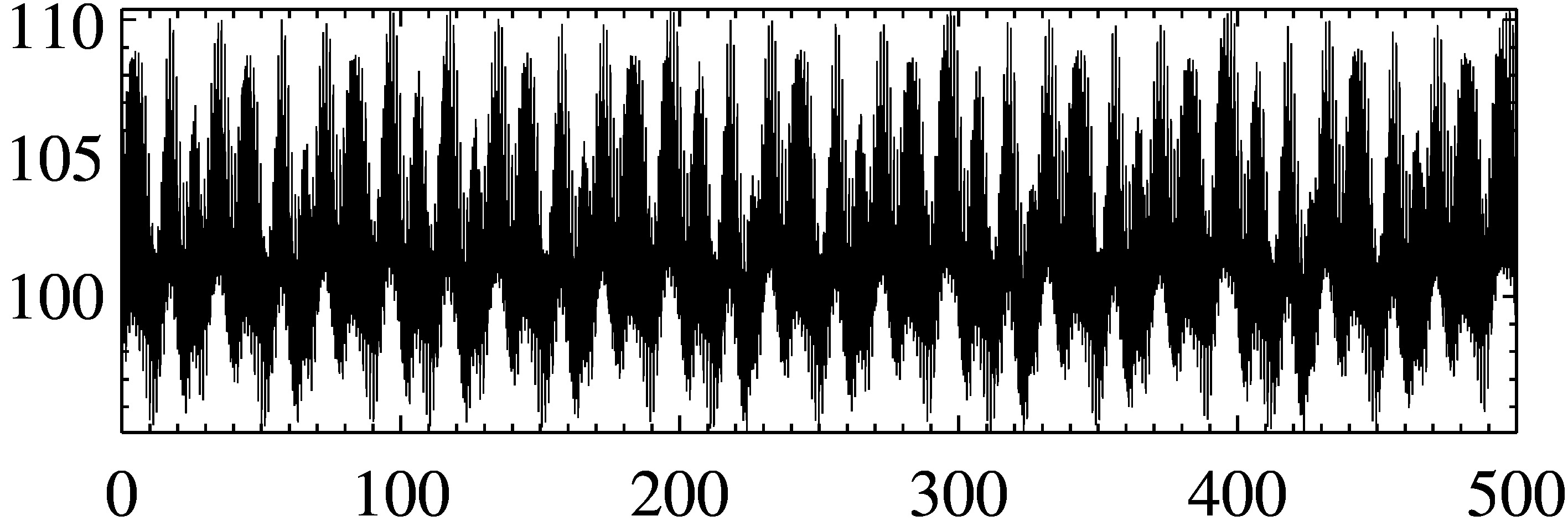}
\includegraphics[scale=0.50]{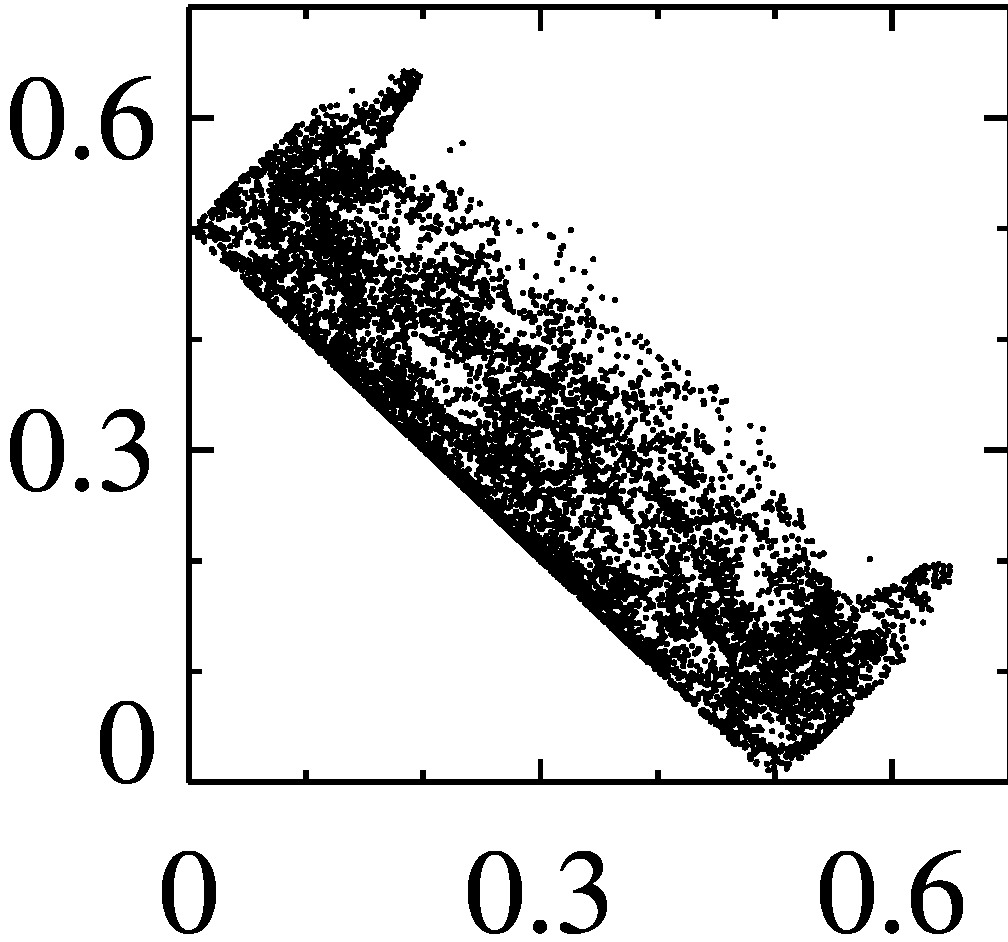}
}
\vspace*{-1.7cm}
\centerline{\hspace*{4.0cm}(c)}
\vspace*{1.2cm}
\centerline{
\includegraphics[scale=0.50]{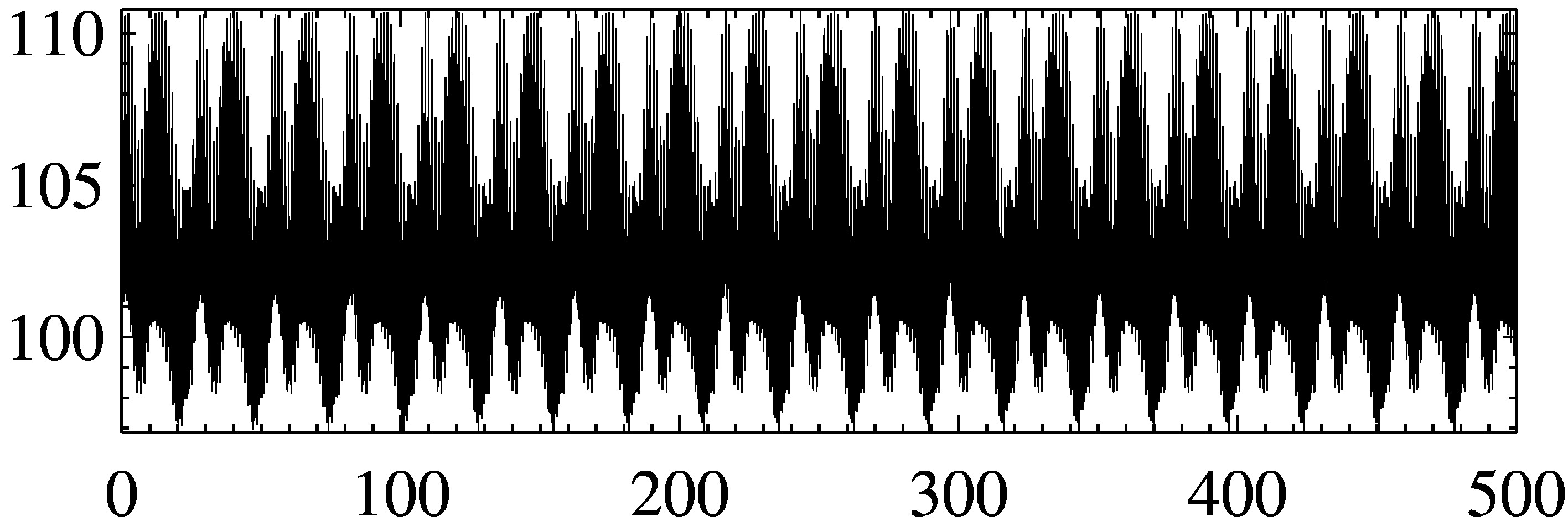}
\includegraphics[scale=0.50]{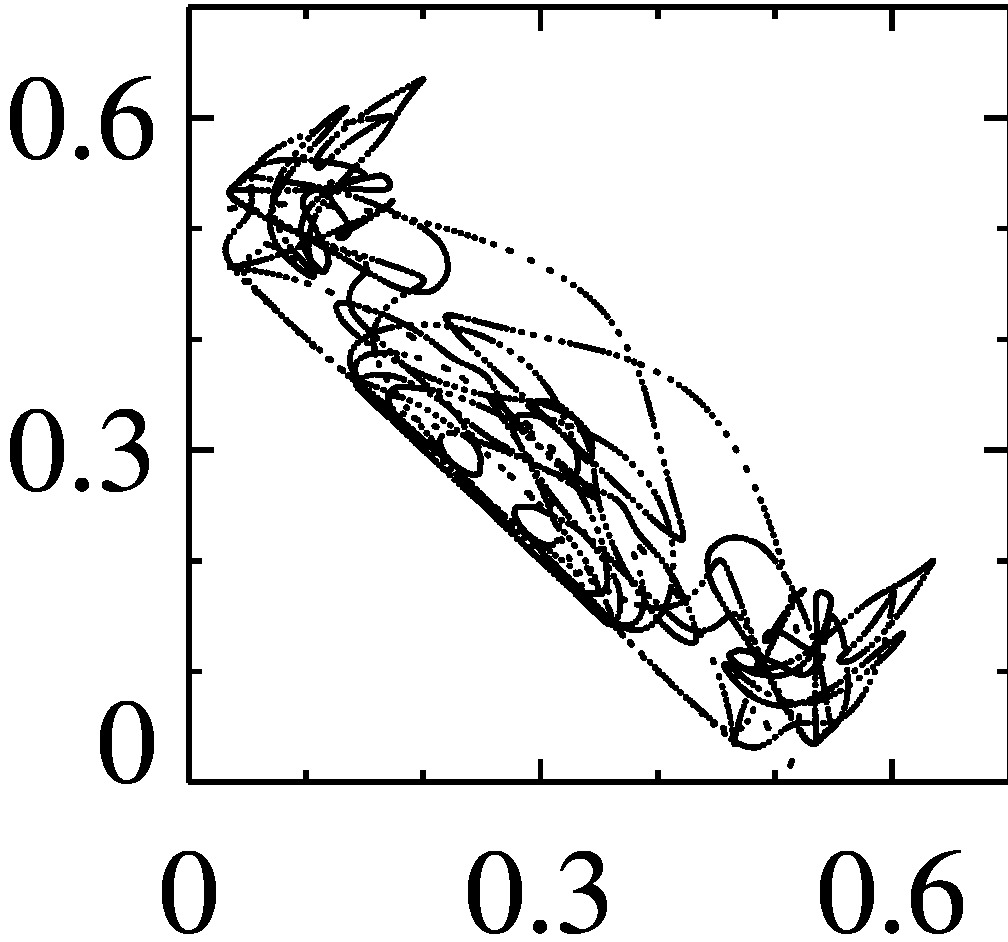}
}
\vspace*{-1.7cm}
\centerline{\hspace*{4.0cm}(d)}
\vspace*{1.2cm}
\centerline{
\includegraphics[scale=0.50]{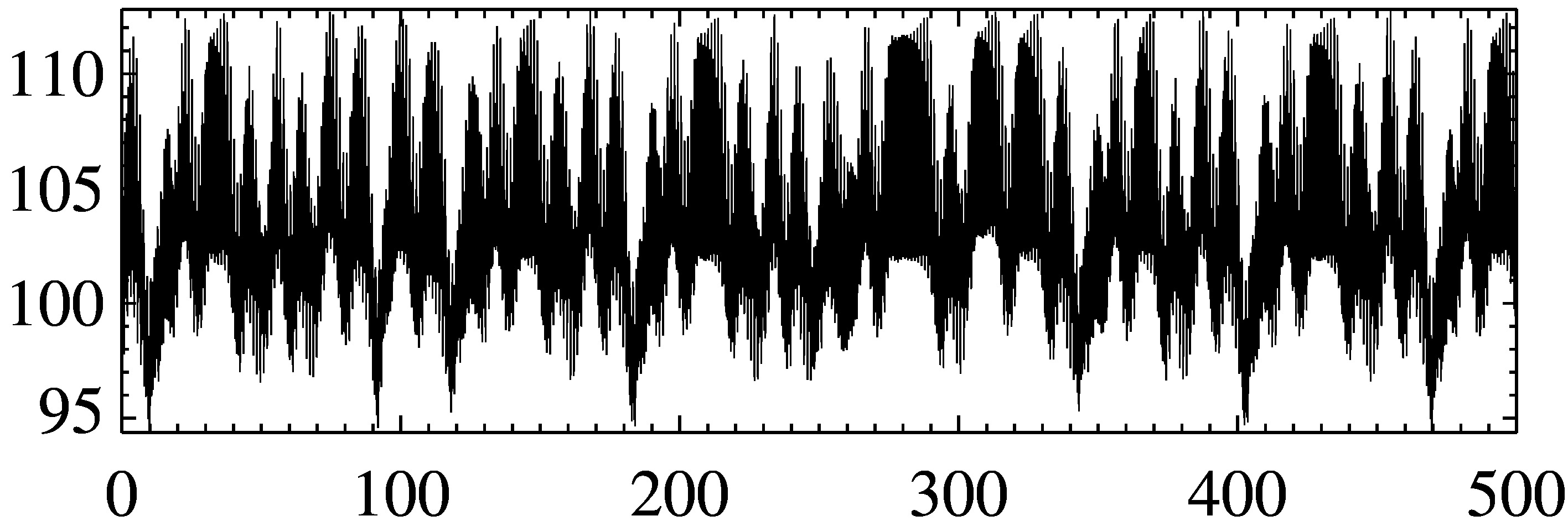}
\includegraphics[scale=0.50]{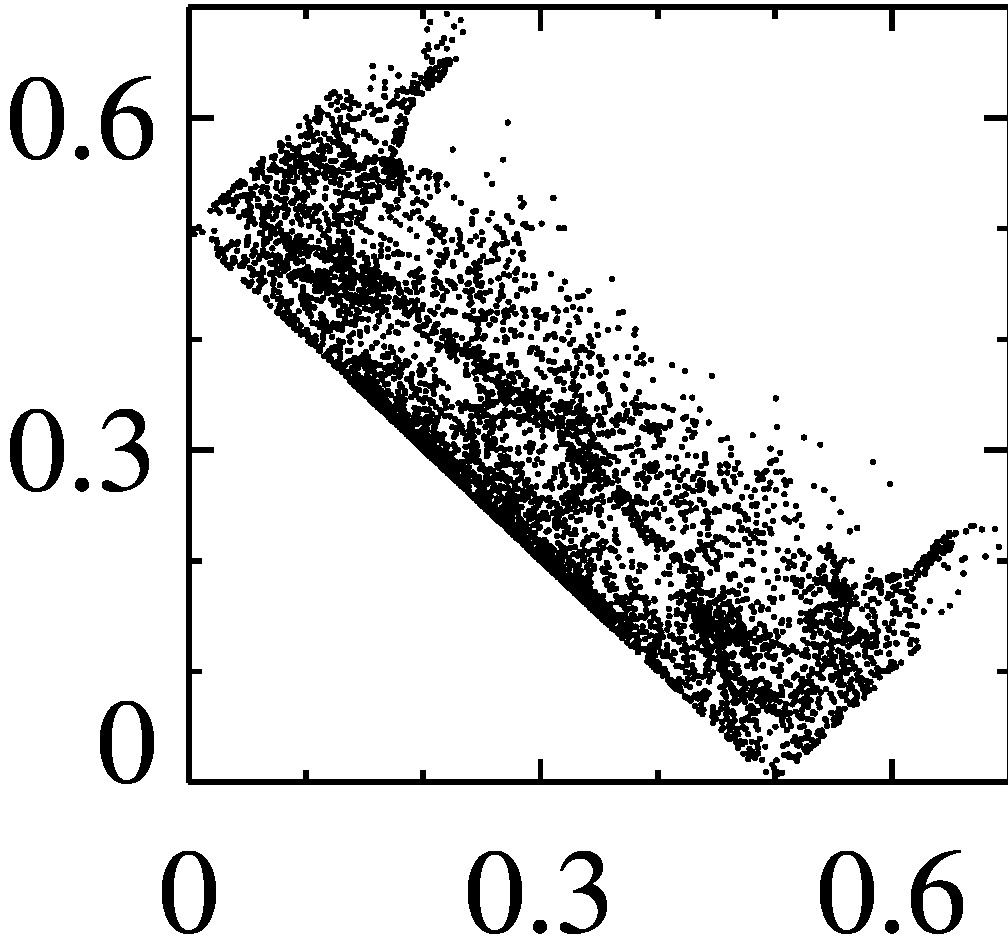}
}
\vspace*{-1.7cm}
\centerline{\hspace*{4.0cm}(e)}
\vspace*{1.2cm}
\centerline{
\includegraphics[scale=0.50]{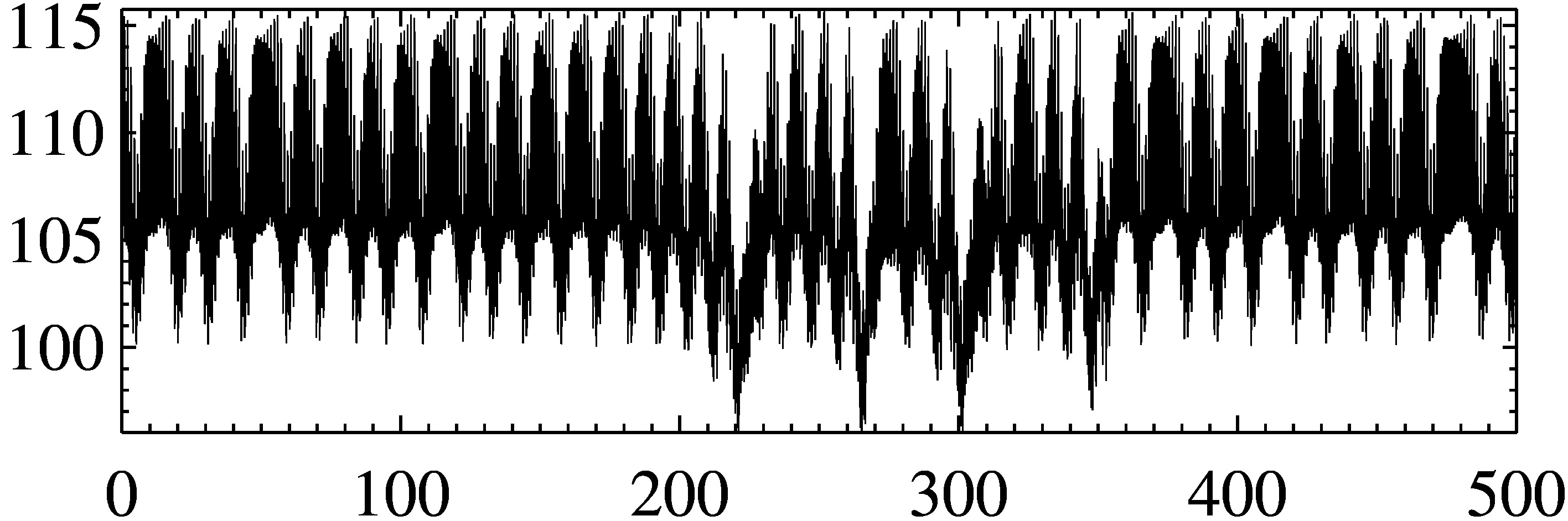}
\includegraphics[scale=0.50]{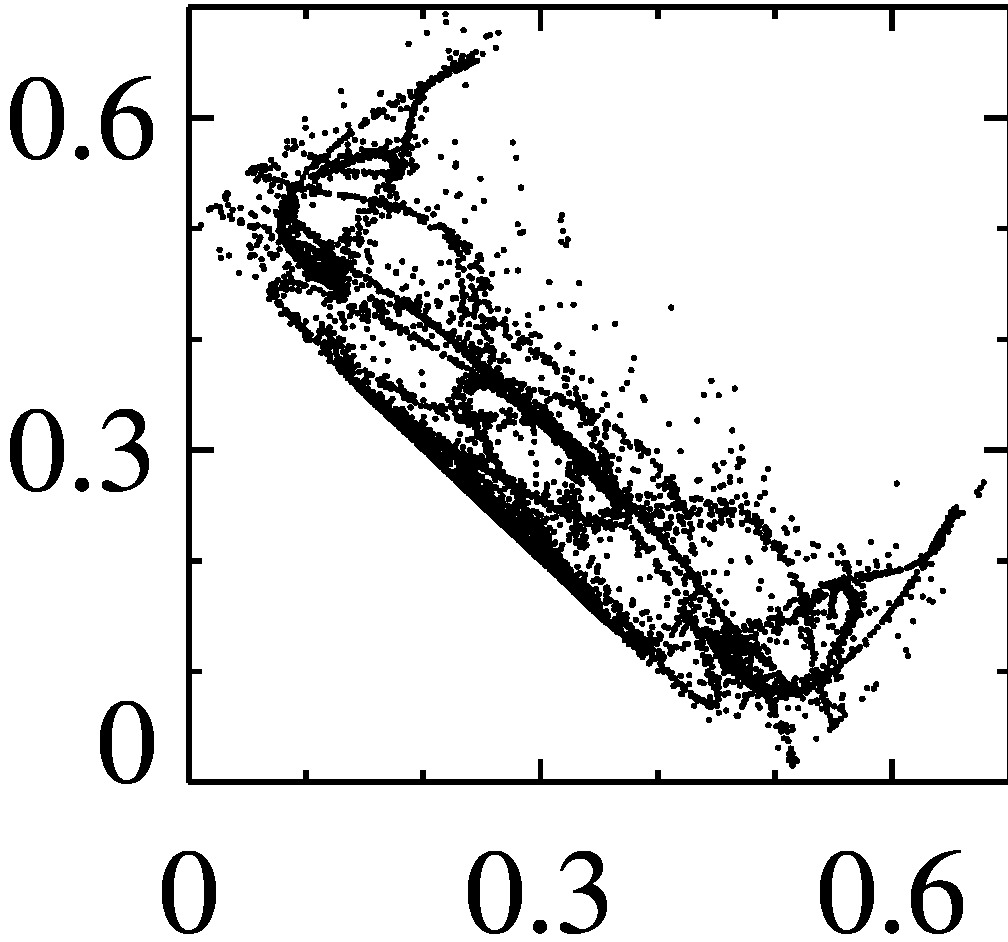}
}
\vspace*{-1.7cm}
\centerline{\hspace*{4.0cm}(f)}
\vspace*{1.2cm}
\centerline{
\includegraphics[scale=0.50]{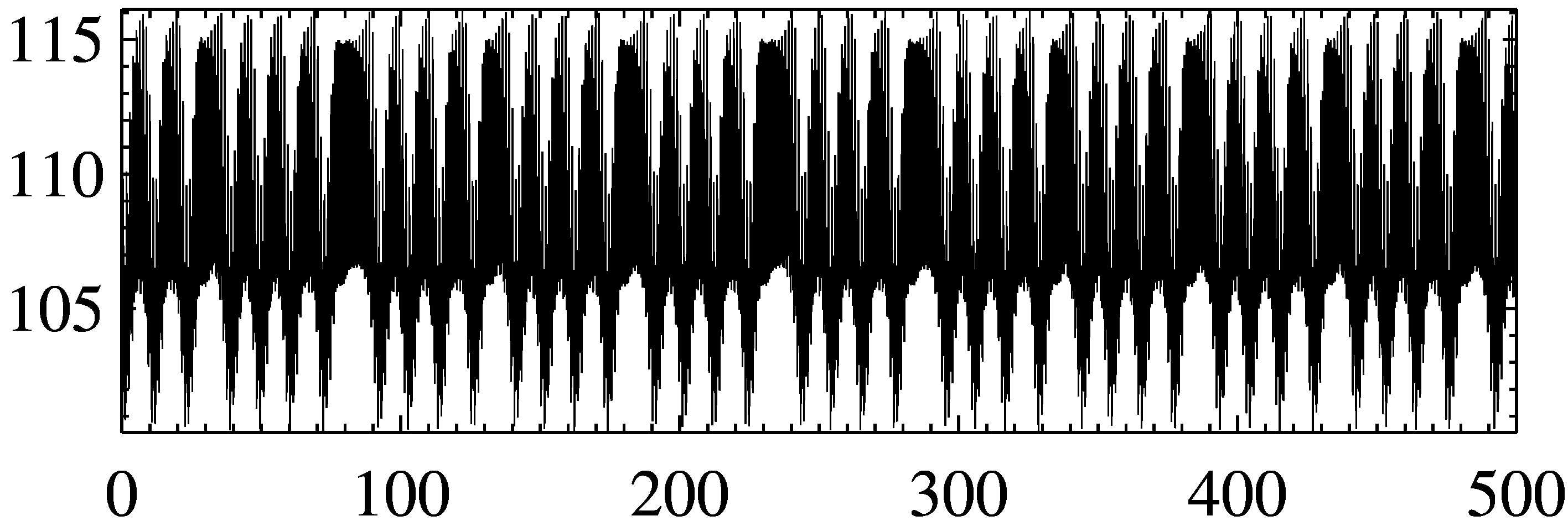}
\includegraphics[scale=0.50]{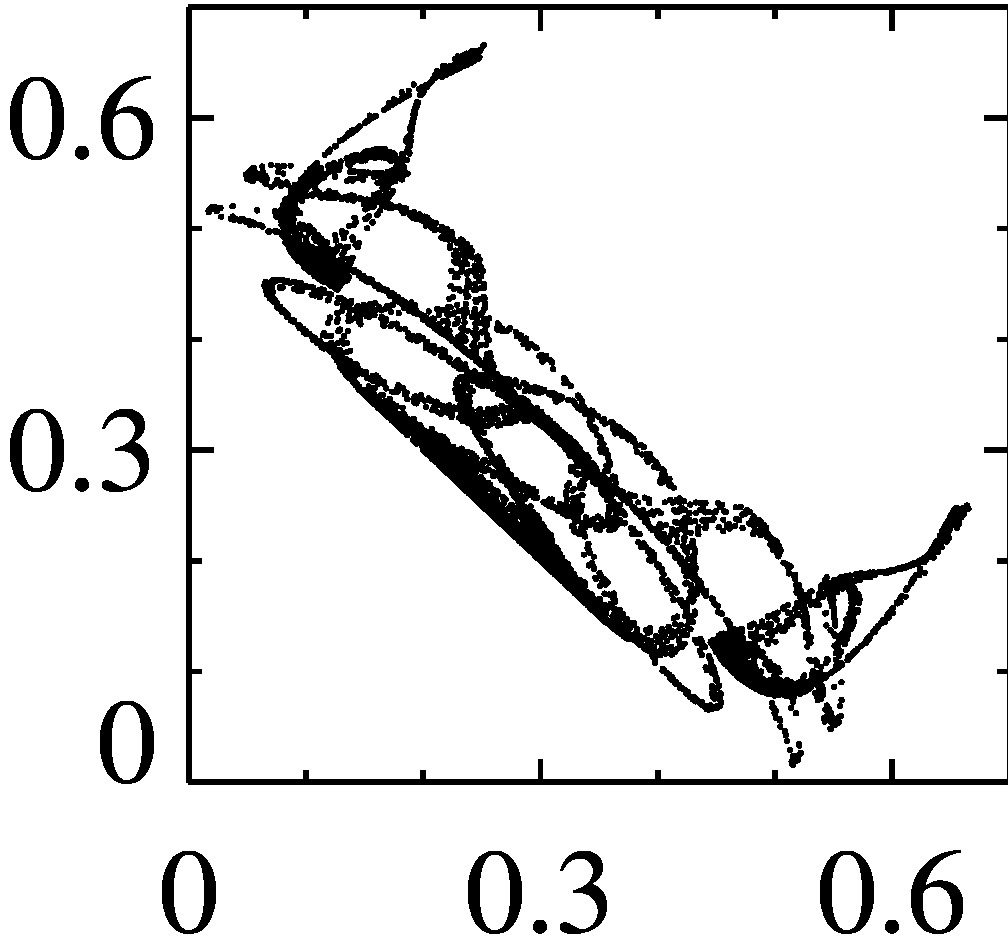}
}
\vspace*{-1.7cm}
\centerline{\hspace*{4.0cm}(g)}
\vspace*{1.2cm}
\centerline{
\includegraphics[scale=0.50]{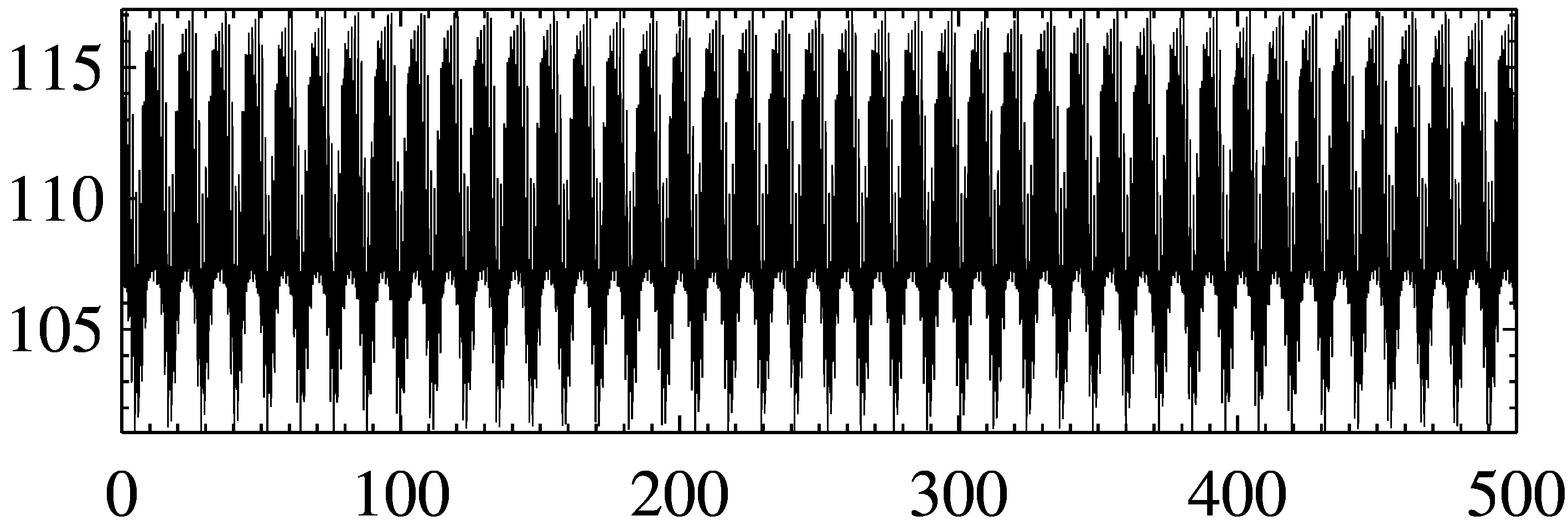}
\includegraphics[scale=0.50]{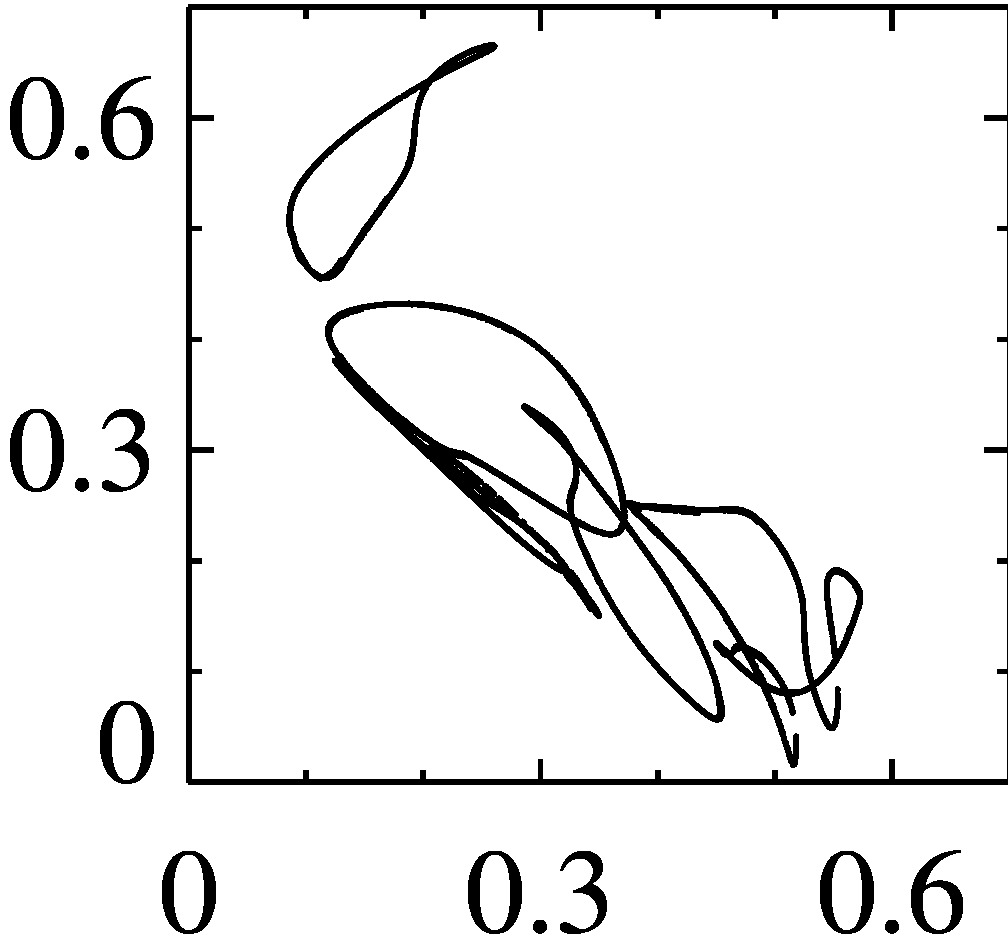}
}
\vspace*{-1.7cm}
\centerline{\hspace*{4.0cm}(h)}
\vspace*{1.2cm}
\centerline{
\includegraphics[scale=0.50]{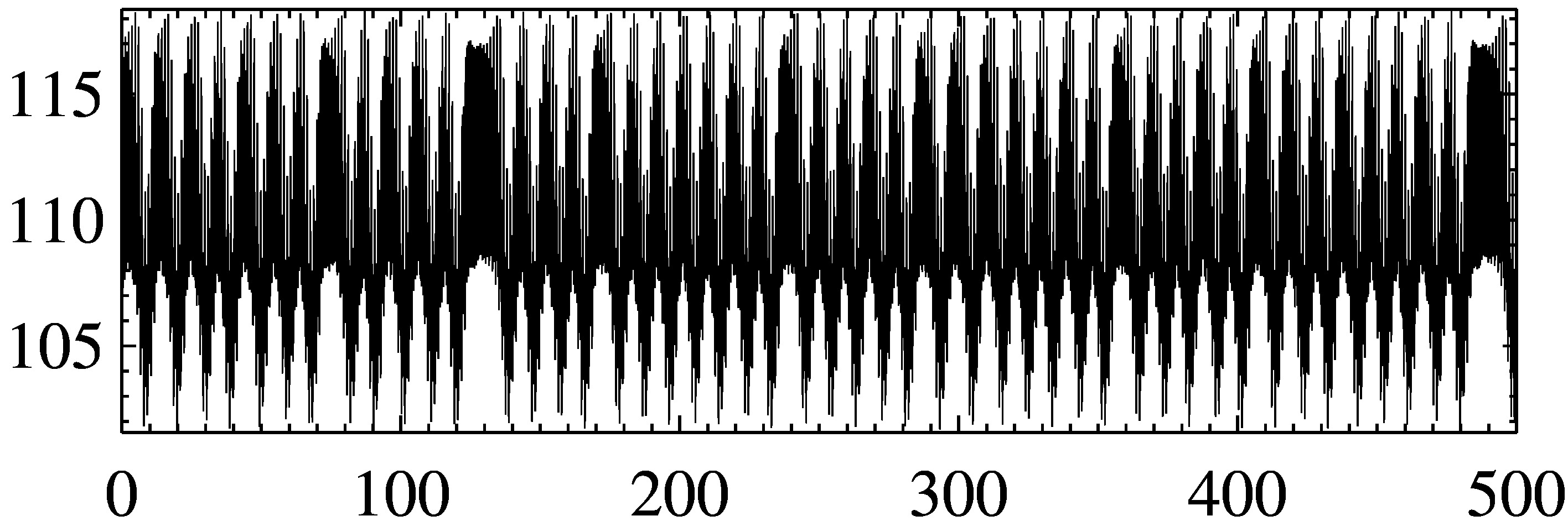}
\includegraphics[scale=0.50]{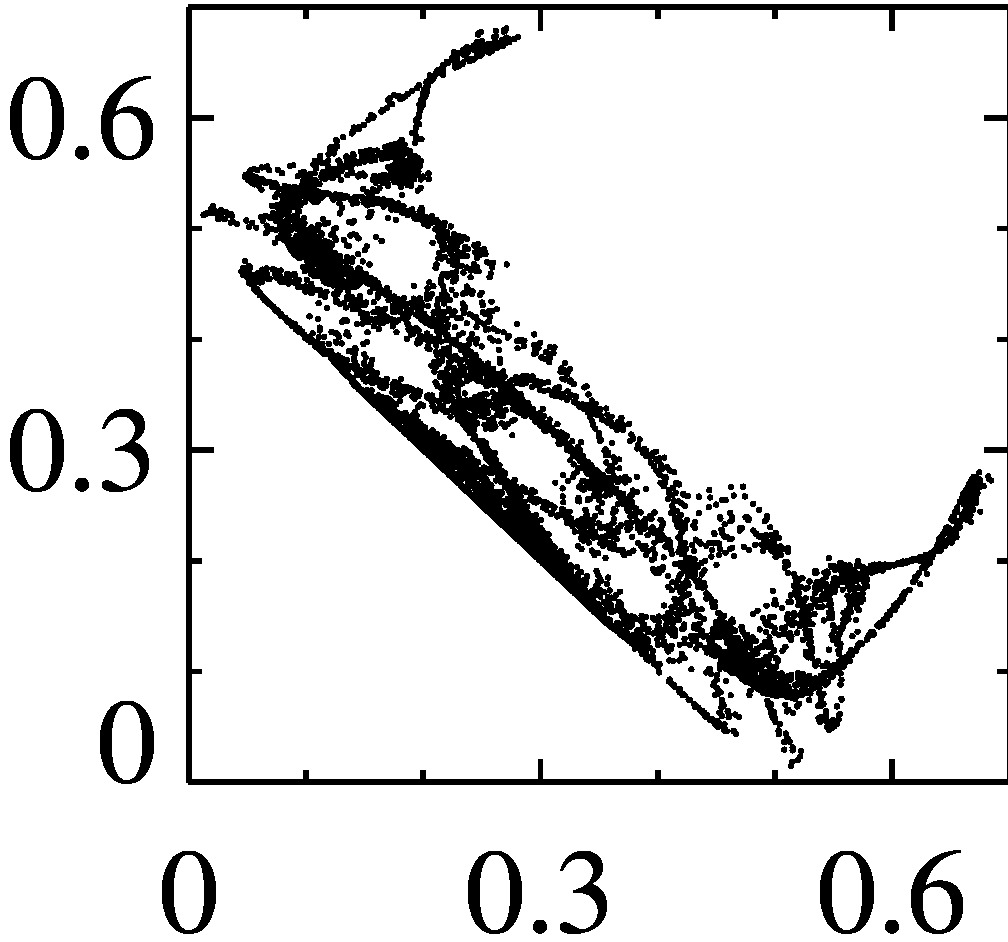}
}
\vspace*{-1.7cm}
\centerline{\hspace*{4.0cm}(i)}
\vspace*{1.2cm}
\centerline{
\includegraphics[scale=0.50]{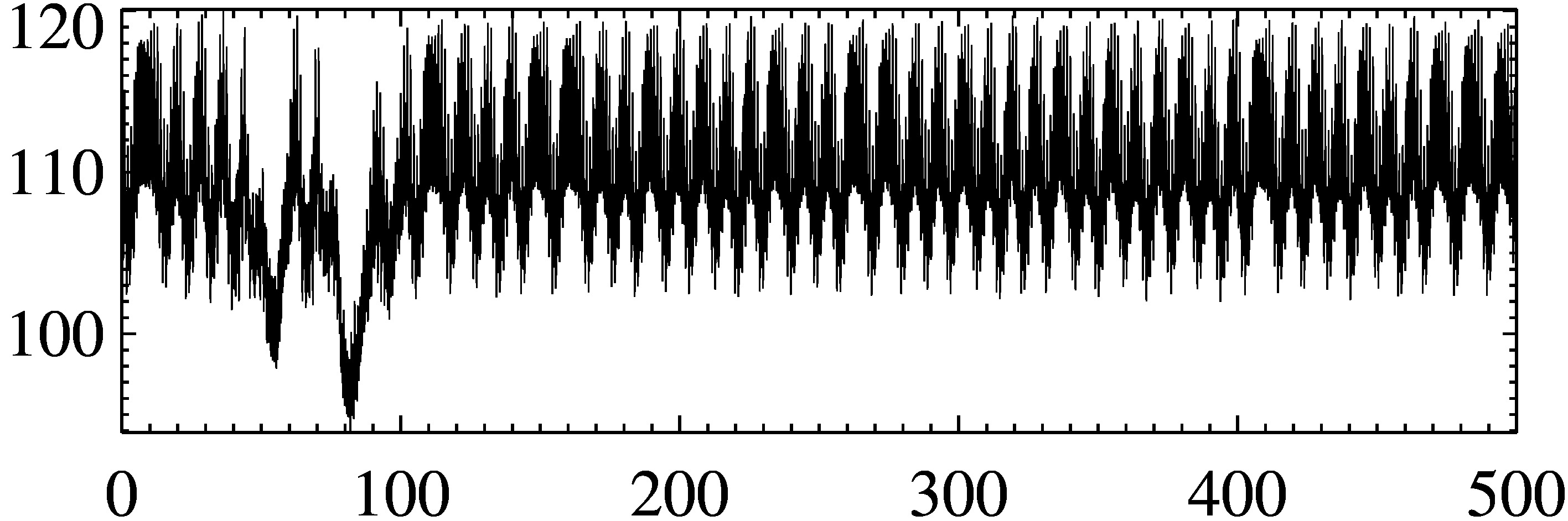}
\includegraphics[scale=0.50]{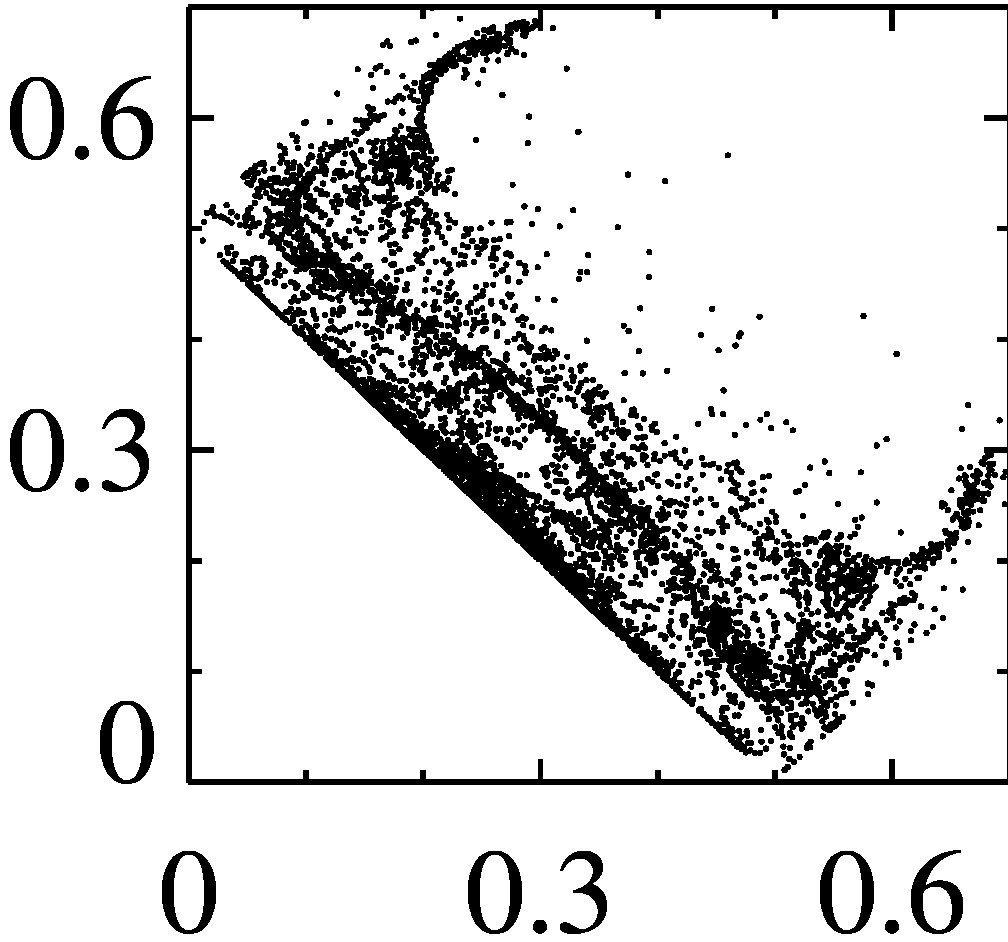}
}
\vspace*{-1.7cm}
\centerline{\hspace*{4.0cm}(j)}
\vspace*{1.2cm}
\caption{ 
Kinetic energy evolution 
and the Poincar\'e section for 
the attractors of the family B$_1$ for
Ra=1880~(a), 
Ra=1910~(b), 
Ra=1990~(c), 
Ra=2000~(d), 
Ra=2010~(e),
Ra=2035~(f),
Ra=2040~(g),
Ra=2050~(h),
Ra=2060~(i),
Ra=2070~(j).
The axes are as in fig.~\ref{fig:tu}. 
}
\label{fig:po}
\end{figure}

Coexisting with $\rm{A}_2$, there is a family of quasiperiodic and
chaotic attractors denoted by $\rm{B}_1$ which, on increasing Ra, gains stability at
Ra=1880. For $1880\le{\rm Ra}\le1900$ the attractors in this family are quasiperiodic with
two basic (incommensurate) frequencies (see fig.~\ref{fig:po} (a)). On 
increasing Ra a period doubling bifurcation occurs, cf.
fig.~\ref{fig:po} (a) and (b), whereby the lowest frequency is halved; this
quasiperiodic attractor exists in $1910\le{\rm Ra}\le1980$. As Ra is
increased further, a sequence of chaotic and quasiperiodic regimes is
observed: for Ra=1990 the convective attractor is chaotic
(fig.~\ref{fig:po} (c)); for Ra=2000 it is quasiperiodic
(fig.~\ref{fig:po} (d)); for $2010\le{\rm Ra}\le2040$ regimes are
chaotic; for Ra=2050 they are quasiperiodic (fig.~\ref{fig:po} (h)) and
finally, for $2060\le{\rm Ra}\le2070$ they are chaotic
(fig.~\ref{fig:po} (i) and (j)). Measurements of the three largest
Lyapunov exponents reveal that all chaotic attractors in this sequence
have one positive and two small in modulus Lyapunov exponents.
Although the identification of each bifurcation occurring in this
interval is out of the scope of our study, we note that {\it i}) windows of
quasiperiodicity can be attributed to frequency locking \cite{ott}, 
{\it ii}) some chaotic attractors in the sequence display intermittent
behaviour, i.e., irregular energy bursts (see fig.~\ref{fig:po} (f) for
$200\le{t}\le350$ and fig.~\ref{fig:po} (j) for $40\le{t}\le110$) randomly occur on the relatively
``smooth background'' reminiscent of regimes shown on fig.~\ref{fig:po}
(g) and (i).      

\begin{figure}
\centerline{
\includegraphics[scale=0.52]{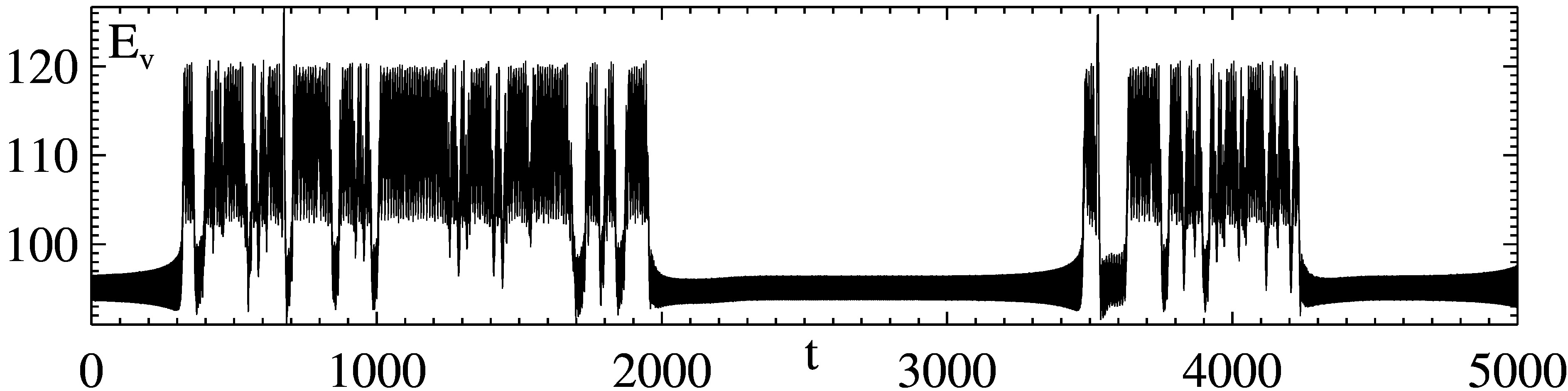}
}
\centerline{
\includegraphics[scale=0.52]{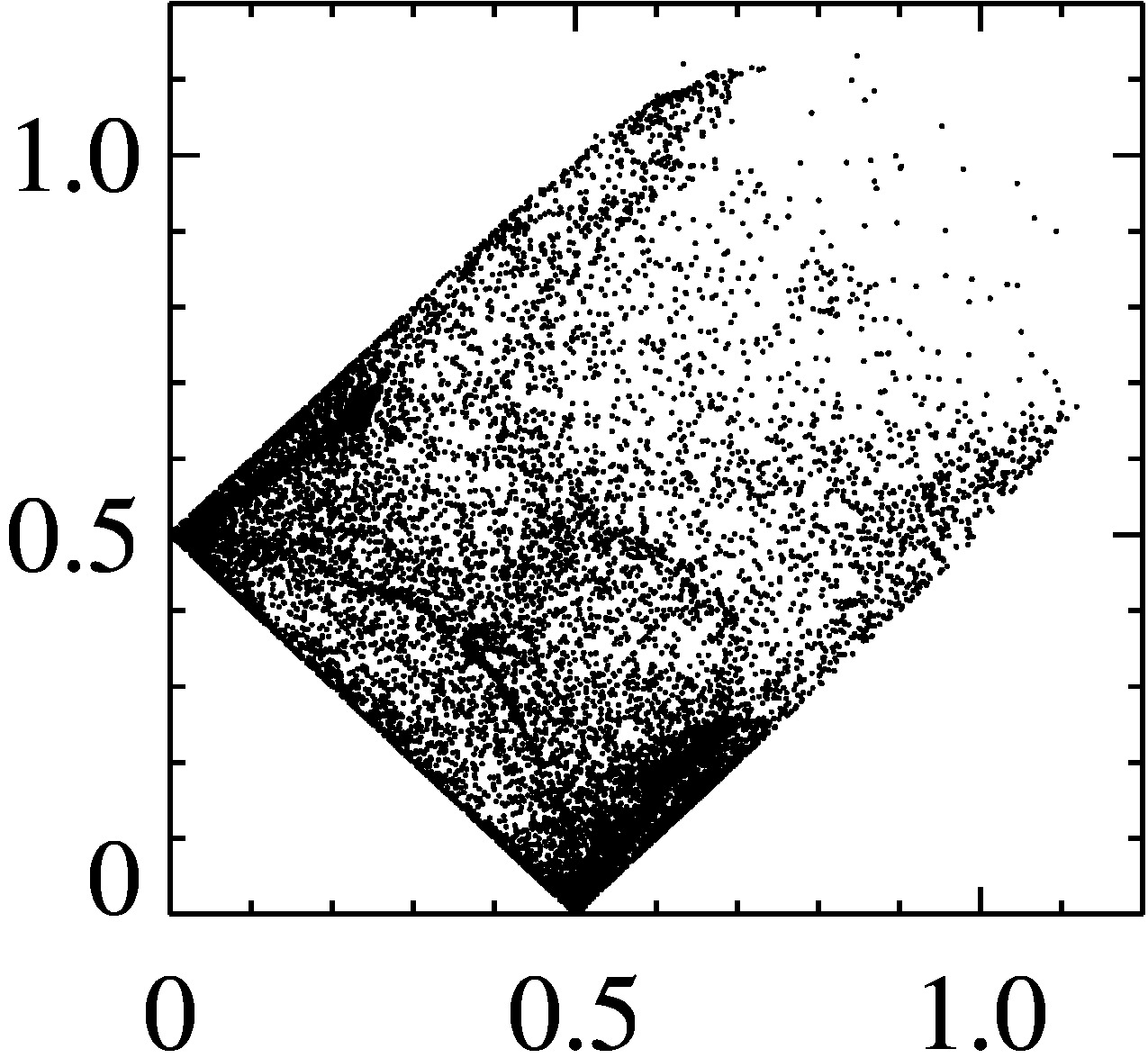} \hspace*{-0.2cm}
\includegraphics[scale=0.52]{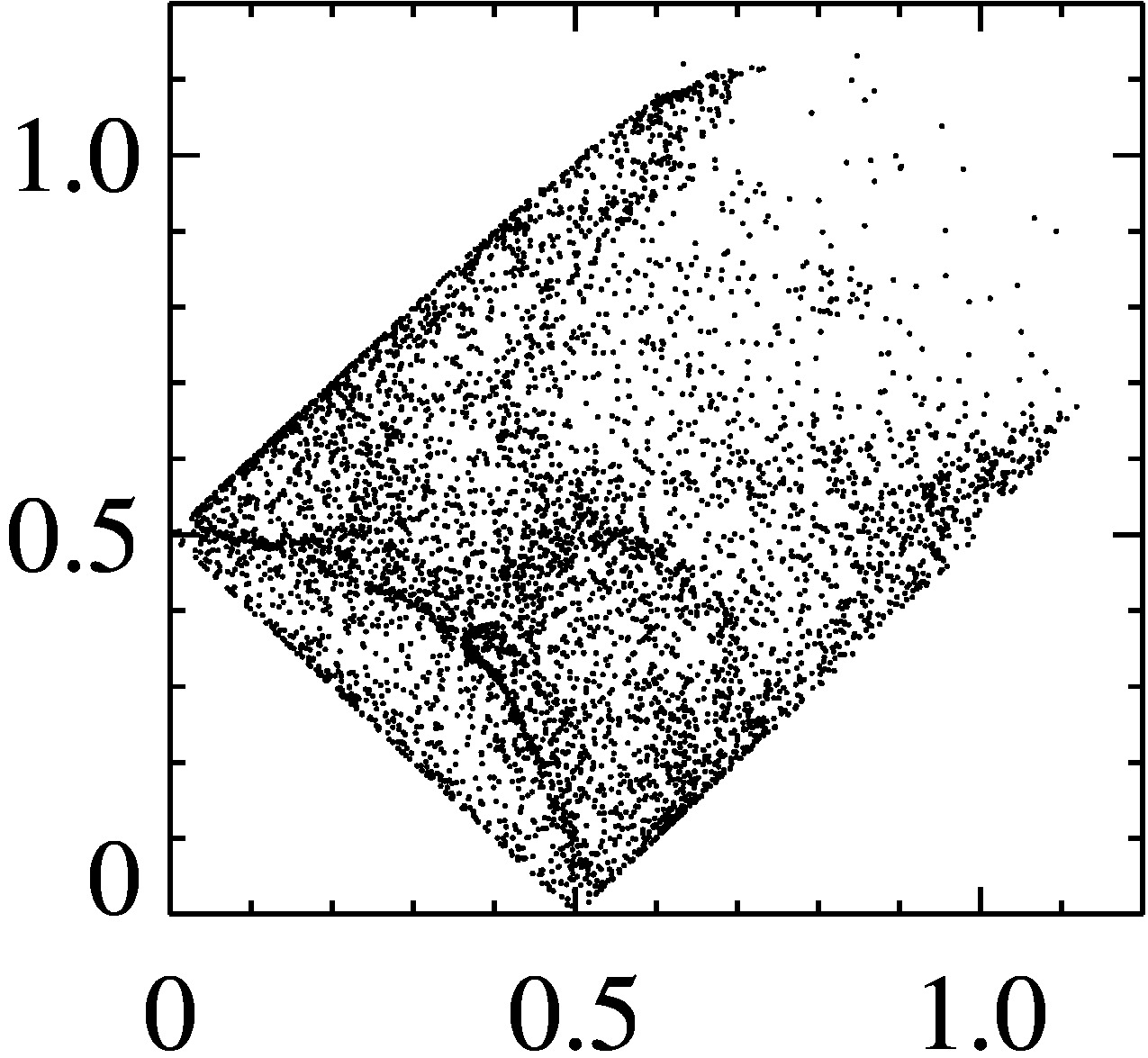} \hspace*{-0.2cm}
\includegraphics[scale=0.52]{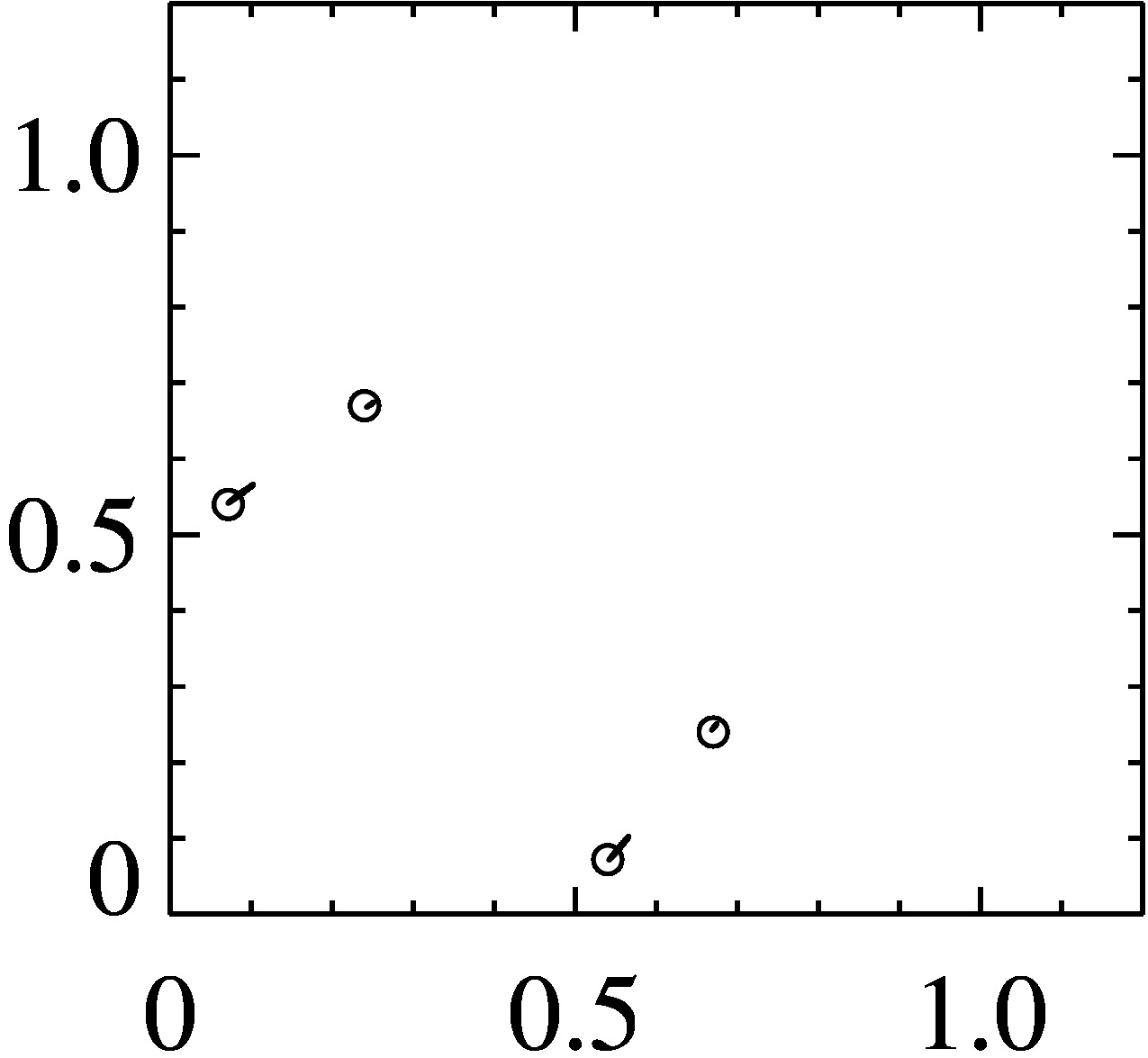}
}
\centerline{
\includegraphics[scale=0.52]{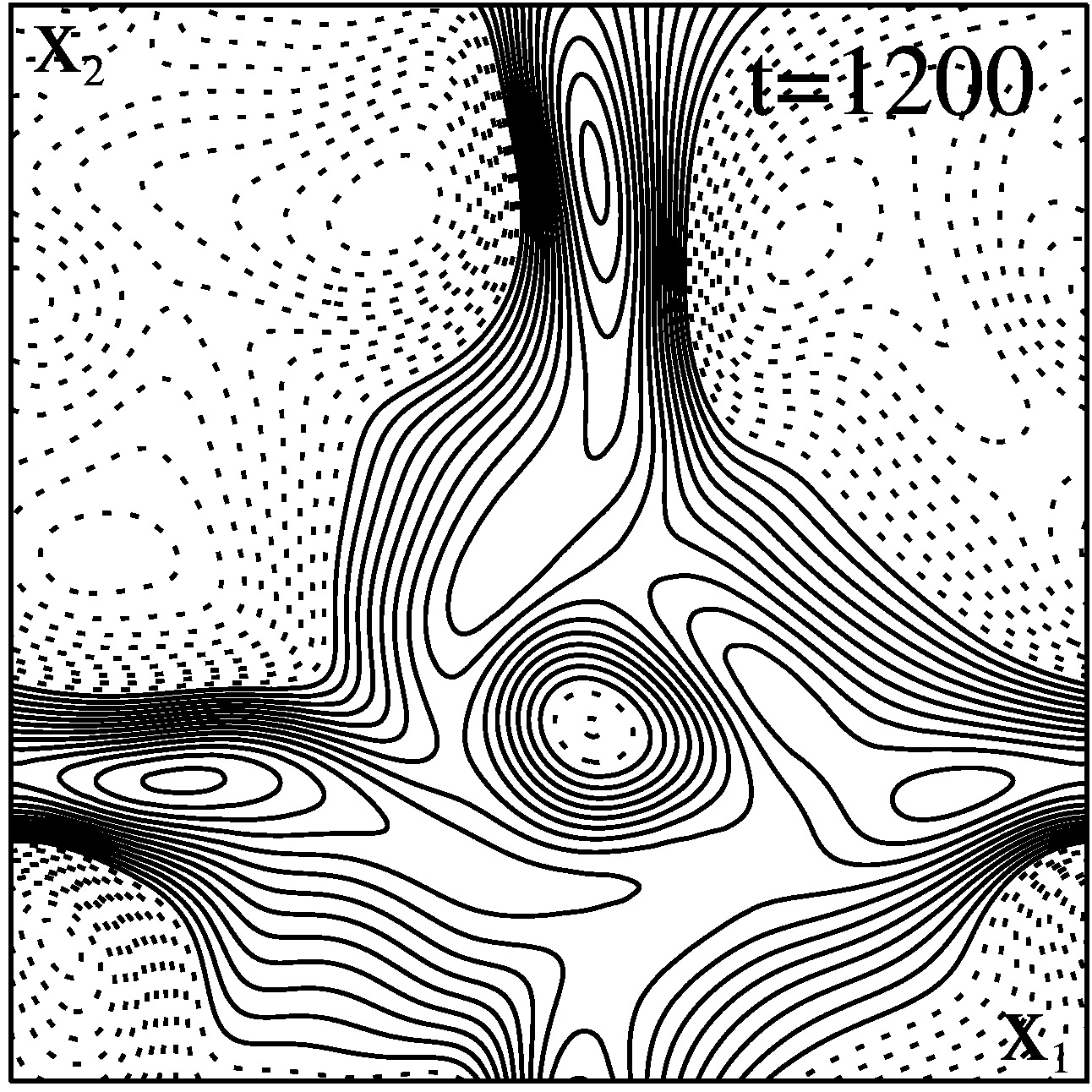} \hspace*{ 0.5cm}
\includegraphics[scale=0.52]{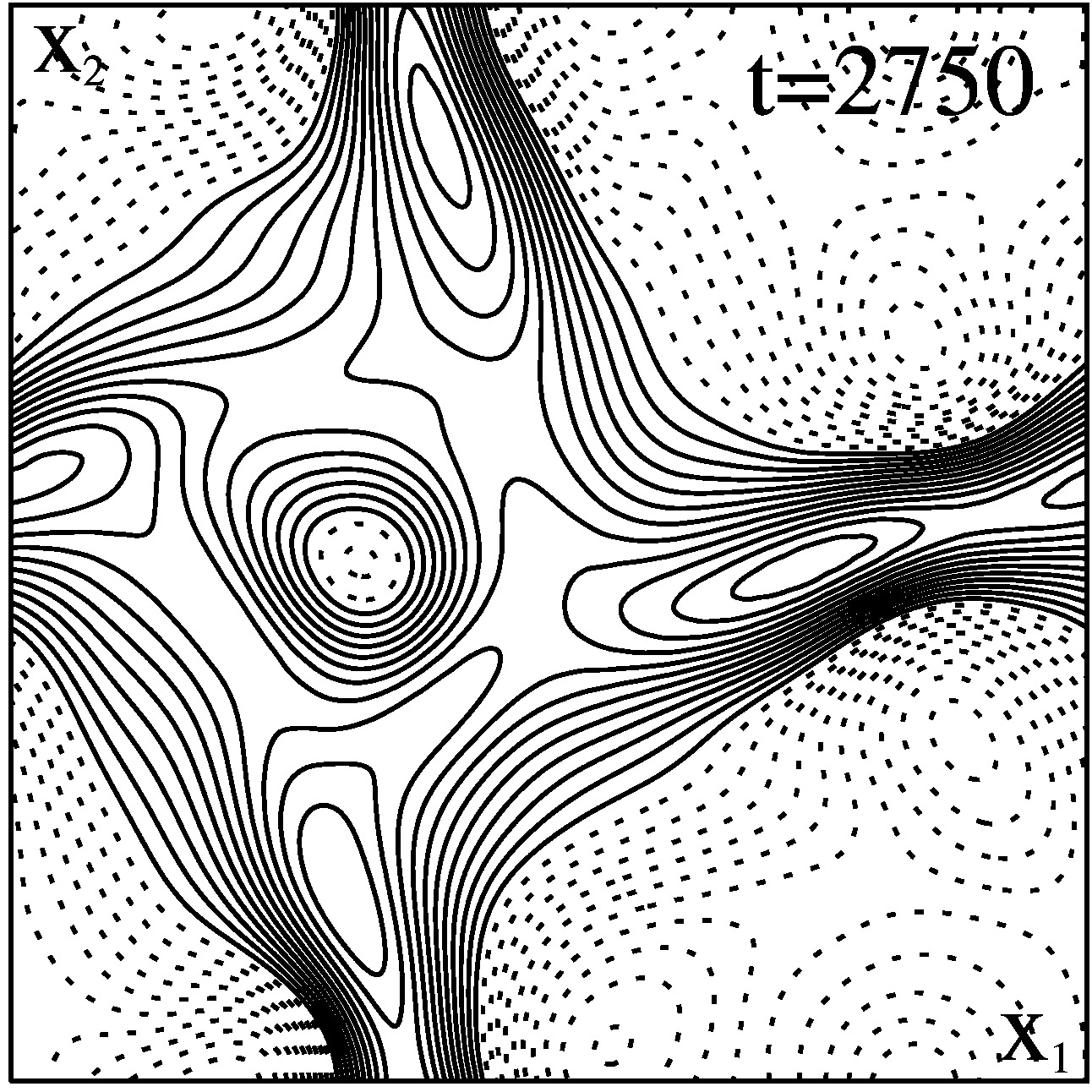}
}
\vspace*{-5.5cm}
\hspace{0.7cm}(a) \hspace{2.3cm}(b) \hspace{2.3cm}(c)

\vspace*{2.3cm}
\hspace{0.7cm}(d) \hspace{3.05cm}(e)
\vspace*{2.2cm}
\caption{
Kinetic energy evolution, the Poincar\'e sections and 
the vertival velocity profiles for the intermittent attractor at Ra=2075. 
On the top and (a) panels the same as
on fig.~\ref{fig:tu} is shown. Points from (a) for the trajectories
intersecting the Poincar\'e plane for $400\le{t}\le1900$ (b) and 
$2500\le{t}\le3000$ (c).  Four open circles in (c) show where the stable
periodic orbit for Ra=2055 from branch $\rm{A}_2$ intersects the
Poincar\'e plane. 
Isolines of $v_3$ on the horizontal mid-plane $x_3=1/2$ are shown for 
$t=1200$ (d) and $t=2750$ (e) (solid and dashed lines stand for positive and 
negative values, respectively).
}
\label{fig:im}
\end{figure}

\begin{figure}[!]
\centerline{
\includegraphics[scale=0.52]{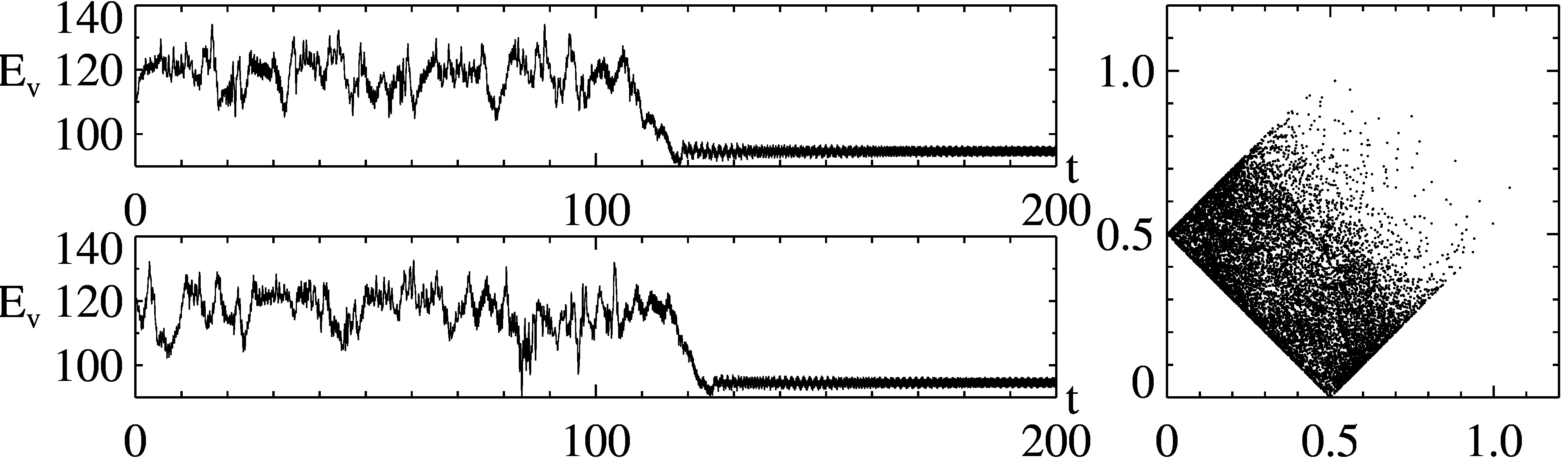}
}
\vspace*{-2.6cm}
\hspace*{5.4cm}(a)\hspace*{0.8cm}(c)

\vspace*{0.9cm}
\hspace*{5.4cm}(b)
\vspace*{0.5cm}

\caption{
Kinetic energy evolution of transients for Ra=2070~((a) and (b)) and 
Poincar\'e section for the transients (c). 
These transients are in the vicinity 
of a chaotic saddle in the phase space. 
The axes are as in fig.~\ref{fig:tu}. 
}
\label{fig:tra}
\end{figure}

For $2073\le{\rm Ra}\le2080$, the family B$_1$ gives rise to an
attractor (B$_2$) displaying ``in-out'' intermittent behaviour \cite{covas01}, 
i.e., the trajectory switches irregularly in time between 
two main states, one is a time-periodic 
state corresponding to the destabilized ${\rm A}_2$, 
the other is chaotic, corresponding to the destabilized 
${\rm B}_1$. Fig.~\ref{fig:im} 
shows the time series, Poincar\'e sections and velocity profiles for the attractor 
at Ra=2075; the trajectory in the
phase space visits the periodic state at $0\le{t}\le200$ and
$2100\le{t}\le3200$. Thus, B$_2$ is formed by a crisis-like event involving the
previously destabilized A$_2$ and B$_1$.  However, a comparison between
the right panel of fig.~\ref{fig:po} (j) and fig.~\ref{fig:im} (a)
reveals that B$_2$ is larger than the union of A$_2$ and B$_1$.  The
missing component can be found from a set of initial conditions
displaying chaotic transients for ${\rm Ra}<2073$. This set constitutes
a chaotic saddle (CS), i.e., a nonattracting chaotic set responsible
for chaotic transients \cite{hsu88,rem04}.
Figs.~\ref{fig:tra}(a) and (b) reveal the kinetic energy time 
series of two initial conditions at
Ra=2070.  Both trajectories exhibit chaotic behavior for about 100 time
units, before they escape from the chaotic region toward the
destabilized A$_2$. Later, the same trajectories will converge to
attractor B$_1$. The Poincar\'e plot for the union of the transient
parts of both series is shown on fig.~\ref{fig:tra} (c) and resembles
the Poincar\'e plots of figs.~\ref{fig:im} (b) and (c). From all this
information, we conjecture the following scenario. 
Attractor A$_2$ is destabilized in a boundary crisis at Ra$\approx$2055 after a
collision with CS, a chaotic saddle lying on the basin boundary between
A$_2$ and B$_1$. The unstable set formed by the union of CS and the
destabilized A$_2$, then, collides with B$_1$ at an interior crisis 
at Ra$\approx$2073 (IC1 in fig.~\ref{fig:en}), leading to the intermittent attractor B$_2$, 
where trajectories alternate between phases where A$_2$, CS and B$_1$ 
can be identified. The highest energy bursts in fig.~\ref{fig:im} 
correspond to CS.

\begin{figure}[!t]
\centerline{
\includegraphics[scale=0.52]{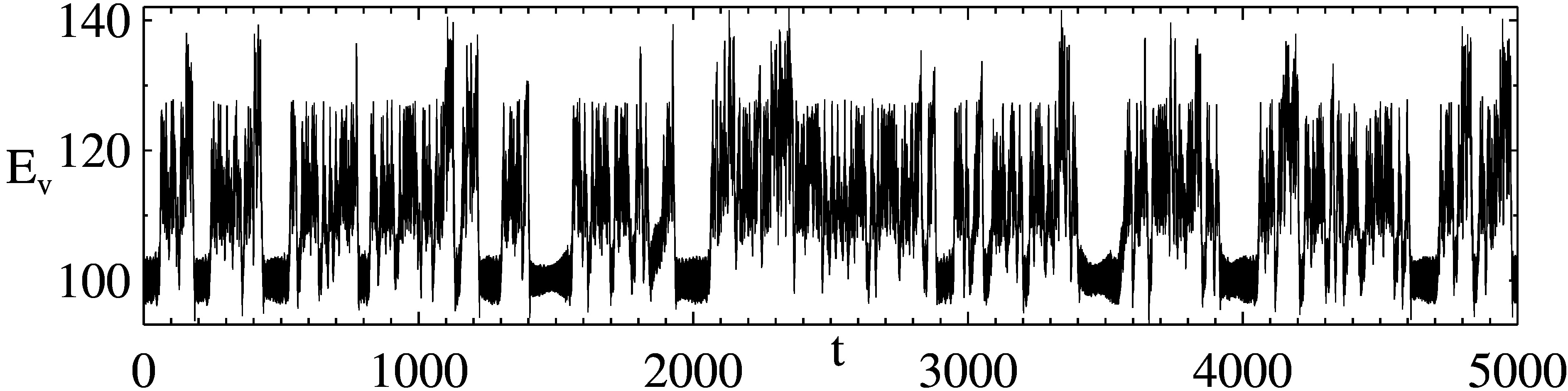}
}
\centerline{
\includegraphics[scale=0.52]{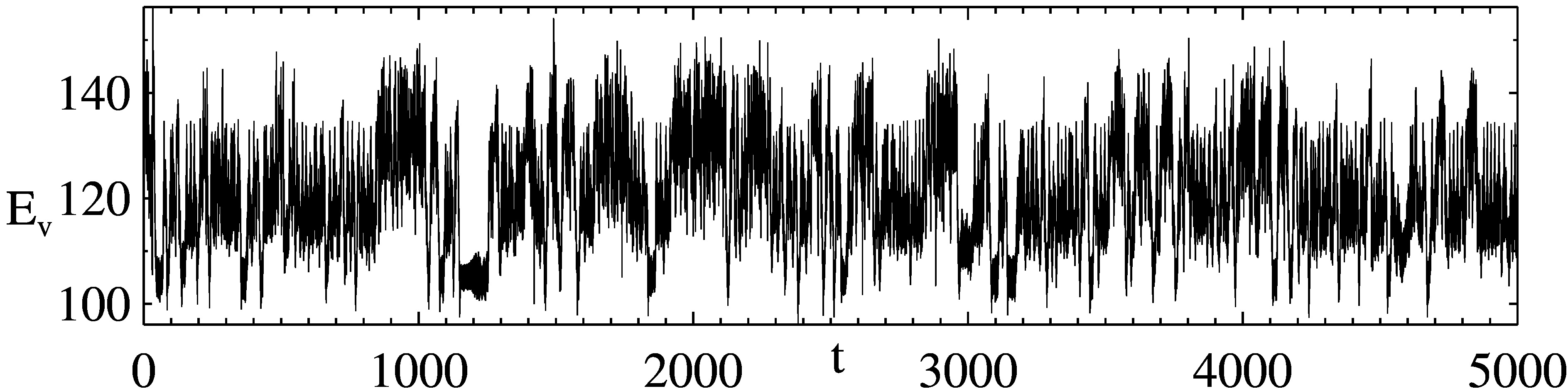}
}
\caption{
Kinetic energy evolution for the attractors of the family B$_3$ 
for Ra=2140 (top panel) and Ra=2200 (bottom panel). 
}
\label{fig:chaim}
\end{figure}

On increasing Ra, the intermittent attractor loses its stability, and 
trajectories for $2090\le{\rm Ra}\le2130$ are
attracted to A$_3$.  However, chaotic transients reminiscent of the
intermittent attractor B$_2$ remain all throughout this interval as a
signature of a chaotic saddle in the background of A$_3$.  
For ${\rm Ra}\ge2140$, the stability of A$_3$ is lost, and a new chaotic attractor,
B$_3$, rises as (apparently) the sole attractor in the system. This attractor
resembles B$_2$ and the chaotic saddle (cf. the kinetic energy evolution for
Ra=2075 from B$_2$ in fig.~\ref{fig:im}, and for Ra=2140 and Ra=2200
from B$_3$ in fig.~\ref{fig:chaim}). In the intermittent regimes of
B$_3$, the time spent near a state with regular behaviour is shortening for
increasing Ra (cf.regimes at Ra=2140, top panel of
fig.~\ref{fig:chaim}, and at Ra=2200, bottom panel of
fig.~\ref{fig:chaim}). Note that some of the regular phases of the
intermittency are in the region previously occupied by A$_3$. This
corroborates another interior crisis scenario 
(IC2 in fig~\ref{fig:en}), whereby attractor A$_3$
collides with the background chaotic saddle to form an enlarged chaotic
attractor B$_3$, where intermittent switches between the former A$_3$
and the chaotic saddle take place.
Such crisis-induced intermittency involving a destabilized attractor and a 
surrounding chaotic saddle was reported for a one-dimensional spatiotemporally chaotic
system in \cite{rem07}.

\begin{figure}[!]
\centerline{
\includegraphics[scale=0.7]{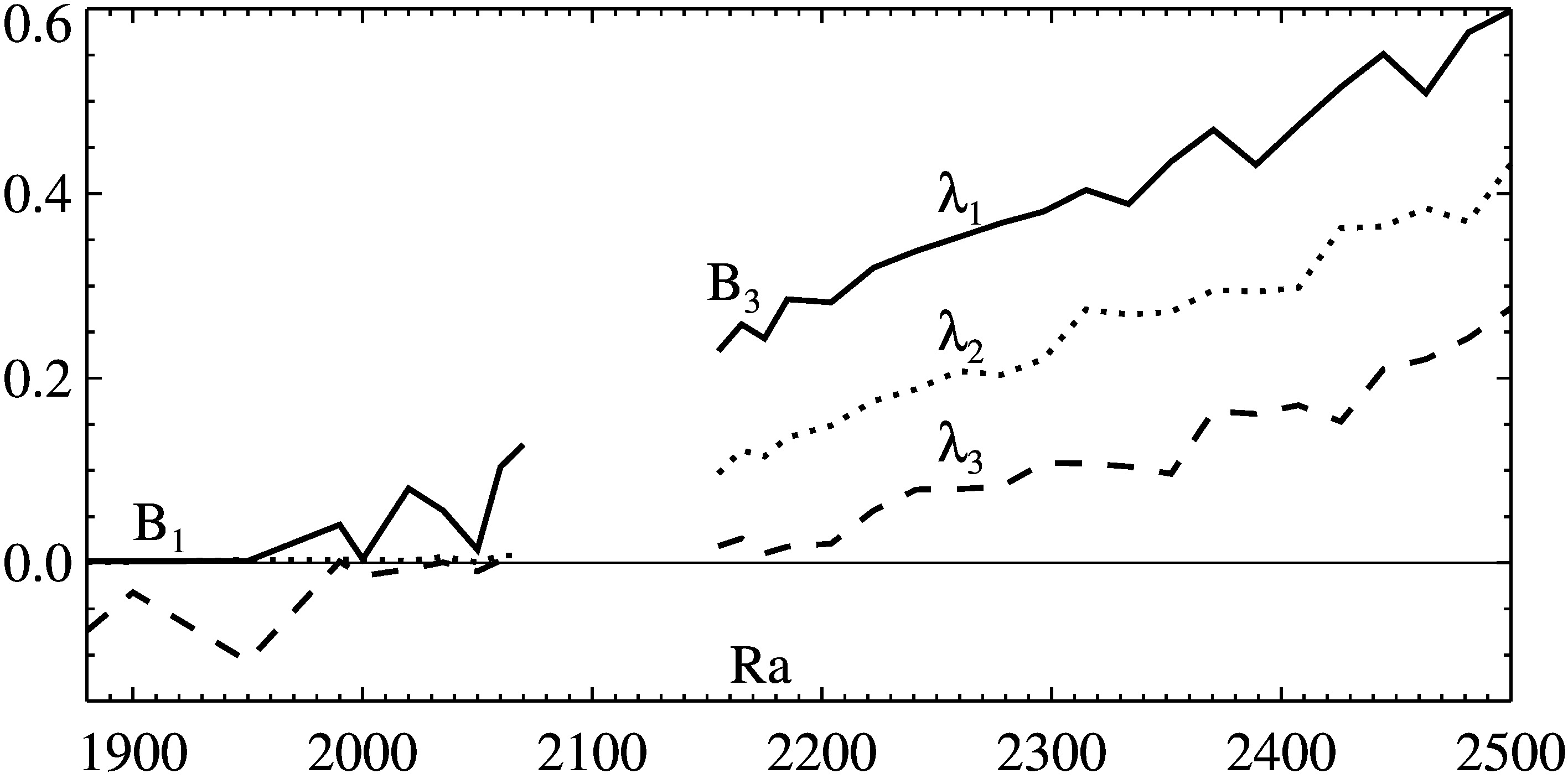}
}
\caption{
The three largest Lyapunov exponents (vertical axis) for the attractors 
of the families B$_1$ and B$_3$. 
}
\label{fig:lya}
\end{figure}

The three largest Lyapunov exponents for the attractors from branches
B$_1$ and B$_3$ of fig.~\ref{fig:en} are shown in fig.~\ref{fig:lya} as
a function of Ra. Hyperchaos with at least three positive Lyapunov
exponents is observed after the interior crisis of B$_3$.  Note the
linear scaling of the exponents with Ra, a feature that has been
observed in Rayleigh-B\'enard convection near the onset for the first
exponent as a function of the reduced Rayleigh number by \cite{jaya06}.
The gap between B$_1$ and B$_3$ is filled by a chaotic saddle that
evolves from the destabilized B$_1$. Although we have not computed the
Lyapunov exponents of the chaotic saddle due to its repelling nature,
we believe it should show a continuation of the linear scaling observed
in fig.~\ref{fig:lya}. 

\section{Conclusions}

We have shown that the evolution from periodic convective states to hyperchaos 
in Rayleigh-B\'enard convection for $P=0.3$ occurs through a sequence of
local and global bifurcations as the Rayleigh number is increased,
including three crises. First, a quasiperiodic attractor A$_2$
coexisting with another attractor B$_1$ collides with a chaotic saddle (CS)
in the basin boundary and loses stability due to a boundary crisis. Evidence
for this is based on A$_2$ becoming transient and the disappearance of its basin of attraction.
Then, the newly formed chaotic saddle (CS + destabilized A$_2$) collides with the chaotic attractor B$_1$ in
an interior crisis (IC1) leading to an enlarged attractor B$_2$. The evidence
is the intermittent switching between B$_1$, CS and A$_2$. Next,
attractor B$_2$ is destabilized and generates a chaotic saddle
surrounding a quasiperiodic attractor A$_3$ (evidenced by B$_2$ becoming transient and
the destruction of its basin of attraction).  Finally, A$_3$ collides
with the surrounding chaotic saddle in another interior crisis (IC2),
which is evident from the intermittent switching between A$_3$ and CS in 
the time series.
The enlarged attractor B$_3$ exhibits hyperchaos with at least three
positive Lyapunov exponents. 

Although strictly speaking a crisis is a collision of a chaotic
attractor with a saddle invariant set \cite{greb87}, in a more general
definition, a crisis can be described as a collision between any
attractor and a saddle-type invariant set \cite{witt97}. In the present
letter, two of the identified crises involve the collision of a
quasiperiodic attractor and a chaotic saddle.

There is surprising similarity between our results and the route to
chaos found in \cite{paul11}, where two-dimensional Rayleigh-B\'enard
convection was studied for $P=6.8$. However, the attractors found {\it
ibid.} are of a limited interest in hydrodynamics, because
two-dimensional convective flows for the considered value of $P$ are
stable only in a narrow window of parameter values
\cite{busbol84,bolbus85}. They are also not useful 
in the dynamo theory, since magnetic field
generation by a planar flow is impossible by virtue of the Zeldovich
antidynamo theorem (see \cite{moffat} for details). 
Note that our results also have some correlation with the bifurcation diagram 
shown in \cite{pal} for zero Prandtl number near the onset of convection,
where stationary, oscilating and chaotic regimes are found, although
no hyperchaos was reported then, probably due to the low values of Ra.
Intermittent and chaotic waves were also found in \cite{thual} for
$P=0.2$ as well as $P=0$, with both free-slip and no-slip boundary
conditions for small aspect ratio geometries. 

Experimental verification of some bifurcations reported here have been
found in the past for Rayleigh-B\'enard convection, such as abrupt
transitions from quasiperiodic to chaotic attractors \cite{swego78}, a
route to chaos via intermittency \cite{berge80}, and quasiperiodic
attractors with three basic frequencies \cite{gobe80}. 

Regarding future works, note that we have not fully characterised all
bifurcations presented and, in particular, the destabilisation of B$_2$
(supposedly another boundary crisis) was not investigated. Additionally,
we do not claim to have found all the attractors present in the range of
Ra studied.  Continuing the line of research started in \cite{podv08,physd,podv06}, we
plan to add rotation and magnetic field, whereby one can study
mechanisms of the convective electromagnetic processes in the liquid
outer core of the Earth.  

\medskip
RC and ELR acknowledge financial support from FAPESP
(Brazil, grants 2013/01242-8 and 2013/22314-7, respectively). ELR also
acknowledges financial support from CNPq (Brazil, grant 305540/2014-9).
EVC acknowledges financial support from CAPES.

\bibliographystyle{eplbib}
\bibliography{eman}

\end{document}